%% file: Time_Delay.tex
\documentclass{article}
\pdfoutput=1
\usepackage{jcappub}
\usepackage{amsmath}
\usepackage{amsfonts}
\usepackage{amssymb}
\usepackage{graphicx}
\usepackage{xcolor}

\usepackage{hyperref}
\hypersetup{colorlinks=true, linkcolor=blue, urlcolor=blue, citecolor=blue, breaklinks=true}
\definecolor{darkred}{RGB}{175,0,0}

\title{Breaking binary formation mechanism degeneracies with gravitational wave clustering}

\author[a,b,c]{Nicola Bellomo,}
\emailAdd{nicola.bellomo@unipd.it}

\author[d,e]{Michele Bosi,}
\emailAdd{mbosi@sissa.it}

\author[f]{Sarah Libanore,}

\author[a,b]{Michele Liguori,}

\author[g,h,i,b]{Michela Mapelli,}

\author[j,a,b]{Federico Semenzato,}

\author[k]{Stefano Torniamenti}

\affiliation[a]{Dipartimento di Fisica e Astronomia G. Galilei, Universit\`a degli Studi di Padova, Via Marzolo 8,
I-35131 Padova, Italy}
\affiliation[b]{INFN, Sezione di Padova, Via Marzolo 8, I-35131, Padova, Italy}
\affiliation[c]{INAF - Osservatorio Astronomico di Padova, Vicolo dell'Osservatorio 5, I-35122 Padova, Italy}
\affiliation[d]{Scuola Internazionale Superiore di Studi Avanzati, Via Bonomea 265, 34136 Trieste, Italy}
\affiliation[e]{Department of Physics, University of Trento, Via Sommarive 14, 38123 Povo (TN), Italy}
\affiliation[f]{Department of Physics, Ben-Gurion University of the Negev, Be'er Sheva 84105, Israel}
\affiliation[g]{Universit\"at Heidelberg, Zentrum f\"ur Astronomie (ZAH), Institut f\"ur Theoretische Astrophysik, Albert Ueberle Str. 2, 69120, Heidelberg, Germany}
\affiliation[h]{Universit\"at Heidelberg, Interdiszipli\"ares Zentrum f\"ur Wissenschaftliches Rechnen, D-69120 Heidelberg, Germany}
\affiliation[i]{Dipartimento di Fisica e Astronomia Galileo Galilei, Università di Padova, Vicolo dell’Osservatorio 3, I–35122 Padova, Italy}
\affiliation[j]{Department of Physics and Astronomy, Johns Hopkins University, 3400 North Charles Street, Baltimore, MD, 21218, USA}
\affiliation[k]{Max-Planck-Institut für Astronomie, Königstuhl 17, 69117, Heidelberg, Germany}

\abstract{
Thanks to the almost 400 gravitational wave events detected, we are currently able to grasp the fundamental features of black hole mass, spin, and distance distributions. 
However, such a fast increase in the precision of the measurements does not necessarily correspond to a better theoretical understanding of gravitational wave sources, especially in current scenarios where the number of free parameters is significantly larger than the number of inferred properties of the black hole population.
In this work, we showcase how the landscape of theoretical models can be chipped away by complementary data-analysis strategies, in particular by studying the statistical properties of gravitational wave anisotropic distribution. 
Specifically, we show how gravitational wave clustering is sensitive to two unique features of each binary formation mechanism: the time-delay distribution and the properties of the binary hosts.
First, we consider a model-agnostic scenario and show the impact that different time-delay distributions have on the gravitational wave bias.
Then, we consider a realistic scenario where gravitational wave events are sourced either by isolated binary evolution or dynamical processes in globular clusters, and study how the gravitational wave bias is unique sensitive to the specific properties of the environment.
In both scenarios, we show how the cross-correlation between galaxy and gravitational wave catalogs is able to distinguish between models with different time delays or with different binary sub-populations originated in specific formation channels.
}

\begin{document}
\maketitle

\section{Introduction}

Gravitational wave (GW) astronomy is transitioning into a data-driven science.
The unprecedented level of sensitivity reached during the fourth observing run has led to the observation of about 400 resolved GW events, three times as many as those observed in the first three runs combined~\cite{abac:gwtc4, abac:gwtc5}.
Similarly, an increase in the accuracy of the GW waveform modeling is allowing a better characterization of the intrinsic properties of compact object binaries, especially binary black holes (BBHs).

However, every new release of the GW transient catalog highlights how the interpretation of the progenitor black-hole (BH) population properties still depends on which events are included and which tools are used to analyze current data.
This situation is not unexpected, since the ultimate blessing and curse of Bayesian analysis resides in the freedom to arbitrarily select theoretical priors, i.e., which is the underlying fiducial model used for data analysis.
Although the Bayesian framework provides a robust methodology to deal with this aspect, some unresolved ambiguity remains when trying to perform model selection~\cite{trotta:bayesianstatistics}.
Since the relative probability of different models~$M_j$ in explaining the same observation~$d$, i.e.,~$p(M_1|d)/p(M_2|d)$, depends not only on the Bayes factor~$B_{12}=p(d|M_1)/p(d|M_2)$, but also on the relative probability of the models themselves,~$p(M_1)/p(M_2)$, it becomes almost unavoidable to inject our theoretical expectation or preference into the statistical analysis.

In this work, we explore how to improve our model selection capabilities in the context of GW astronomy. 
In particular, we focus on understanding the BBH formation mechanism and whether the origin of the binaries is connected to the isolated or dynamical channel.
Isolated BBHs descend from the evolution of stellar binaries in the field.
These can turn into merging BBHs through a variety of mechanisms, e.g., stable~\cite{giacobbo:smt} or unstable mass-transfer, with the latter leading to a common envelope phase~\cite{ivanova:ce}.
The mass of each progenitor BH in these systems is limited to~$\approx 45\ \rm M_\odot$~\cite{giacobbo:mbh}, and their spins are preferentially aligned~\citep{rodriguez:spin1, gerosa:spin}.
Mechanisms alternative to mass transfer have already been proposed, such as chemically homogeneous evolution~\cite{marchant:ch} or evolution of triple systems~\citep{silsbee:triples, antonini:triples}, although the characteristic properties of BBHs may vary across different scenarios~\citep{mapelli:pop, et:bb}.
On the other hand, BHs dynamically pair up in dense environments, such as stellar clusters and galactic nuclei, through binary-single interactions~\cite{samsing:enc}, three-body binary formation~\citep{heggie:enc, atallah:enc}, GW captures~\cite{oleary:gwcap}, and gas driven dynamics~\cite{mckernan:agn}.
Their progenitors are expected to feature a broader mass spectrum and an isotropic spin distribution~\cite{rodriguez:spin2}.

Inferring the relative importance of different formation channels is crucial to overcome our theoretical uncertainties in the binary formation process in different astrophysical environments.
Current GW observations~\cite{abac:gwtc5} are uncovering, with increasing accuracy, features in BBH mass and spin distributions
that are hard to reconcile with a single formation channel and that may be the signature of distinct subpopulations.
For instance, the analysis of the events GW190521~\cite{lvk:gw190521} and GW231123~\cite{lvk:gw231123} returns BH masses that are difficult to justify in the isolated scenario. 
Additionally, the recently detected GW241011 and GW241110~\cite{lvk:gw241011gw241110} events present high primary spins and unequal mass ratios, which could be explained as a result of hierarchical BBH mergers in dense stellar environments.
With the increased sensitivity of next-generation detectors~\cite{et:bb}, the redshift evolution of the merger rate density will also become a major probe of the interplay between different channels.

Each of the current theoretical models has enough freedom and/or associated uncertainty to match the current observations by tweaking different aspects of the underlying physical processes. 
The ultimate question then becomes: {\it How do we address this theoretical uncertainty?}
In this work, we propose a new avenue to investigate in which proportions the isolated and dynamical BBH formation mechanisms contribute to the entire population of events.
In particular, we show how the statistics of the GW anisotropic distribution carry characteristic information about the environment where BBHs form and merge.
We also demonstrate how the binary-to-merger time-delay distribution of GW events, which is also a clear signature of the formation channel itself, can have a large impact on determining the magnitude of GW clustering. 
Since GWs also act as a cosmological tracer of the Large-Scale Structure (LSS) of the Universe, upcoming data will lead us to an unprecedented position where we can statistically probe the BBH formation mechanism.

Cross-correlating GW and galaxy data sets allows us to probe the statistical properties of the anisotropic distribution of both tracers at the same time and to overcome the noisiness of GW data.
The scientific potential of this technique has already been widely established in terms of probing multiple facets of the Cosmological Standard Model and its components~\cite{oguri:gwxlss, raccanelli:gwxlssI, raccanelli:gwxlssII, scelfo:gwxlssI, calore:gwxlss, mukherjee:gwxlssI, libanore:gwxlssI, scelfo:gwxlssII, scelfo:gwxlssIII, libanore:gwxlssII, mukherjee:gwxlssII, mukherjee:gwxlssIII, scelfo:gwxlssIV, bosi:gwxlssI, libanore:gwxlssIII, balaudo:gwxlss, afroz:gwxlss, zazzera:gwxlssI, pedrotti:gwxlss, zazzera:gwxlssII, sala:gwxlss, deleo:gwxlss, chakravarti:gwxlss, bosi:gwxlssII, scarpel:gwxlss}.
In particular, the authors of a few of these Refs. have already explored the possibility of distinguishing between astrophysical and primordial BHs using GW clustering.
In this work, we follow that very same strategy, but for different classes of exclusively astrophysical formation mechanisms, since the interplay between different time-delay distributions for the two formation channels and different evolution histories of the two characteristic environments provides a rather unique picture in terms of GW clustering strength.

Specifically, we show how cross-correlations between next-generation GW observatory data sets and current/future galaxy surveys can provide a sufficiently small error on the GW bias to disentangle whether multiple formation channels are at play.
First, we demonstrate in an agnostic scenario that GW$\times$LSS cross-correlations have the potential to infer the properties of the time-delay distribution by accurately measuring the GW bias, even without assuming unique environment properties for the underlying binary formation mechanism.
In particular, we show that the longer the time-delay is, the more massive the GW galaxy hosts are and the larger the GW bias turns out to be, complementing the connection between the time-delay distribution and the cosmic assembly of the host galaxies first shown in Ref.~\cite{Artale:2019tfl} and the preliminary study done in Ref.~\cite{chakravarti:gwxlss}.
Second, we simulate from first principles two populations of BBH that form either via isolated binary evolution or dynamical processes in globular clusters, along with a semi-analytical model to populate a realistic realization of the Universe.
Also in this case, we demonstrate how different formation mechanisms, i.e., the combination of different environments properties and time-delay distributions, are responsible for inherently different GW biases, and that future observatories will be able to constrain the presence of multiple formation channels.

The paper is organized as follows.
Section~\ref{sec:gw_clustering} contains a brief theoretical overview of the theory of GW$\times$LSS cross-correlations.
Section~\ref{sec:binary_formation_channels} shows the potential of this technique for both an agnostic and a theoretically-motivated scenario.
Finally, Section~\ref{sec:outlook_conclusions} contains our conclusions.
Appendices~\ref{app:tracer_population}, \ref{app:synthetic_universe}, \ref{app:BBH_formation} describe the technical details implemented in this work.

%%%%%%%%%%%%%%%%%%%%%%%%%%%%%%%%%%%%%%%%%%%%%%%%%%%%%%%%%%%%%%%%%%%%%%%%%%%%%%%%%%%%%%%%%%%%%%%%%%%%%%%%%%%%%%%%%%%%%%%%%%%%%%%%%%%%

\section{Gravitational wave clustering}
\label{sec:gw_clustering}

%%%%%%%%%%%%%%%%%%%%%%%%%%%%%%%%%%%%%%%%%%%%%%%%%%%%%%%%%%%%%%%%%%%%%%%%%%%%%%%%%%%%%%%%%%%%%%%%%%%%%%%%%%%%%%%%%%%%%%%%%%%%%%%%%%%%

\subsection{GW$\times$LSS cross-correlations}

Gravitational wave events, in analogy to galaxies, are biased tracers of the underlying dark matter distribution.
At large scale, number count fluctuations are Gaussian and the sky cannot be approximated as flat; therefore, for the purpose of studying GW$\times$LSS cross-correlations, we consider only two-point functions in harmonic space, i.e., the angular power spectra.
Given two tracers~$X,Y$ in two redshift bins centered at~$z_i,z_j$, respectively, their observed angular power spectrum reads as
\begin{equation}
    C_{\ell,\mathrm{obs}}^{XY} (z_i,z_j) = C_\ell^{XY} (z_i,z_j) + N^{XY}_\ell (z_i,z_j),
\label{eq:cl_general}
\end{equation}
where~$C_\ell$ and~$N_\ell$ are the angular power spectra of the physical signal and the noise, respectively.
The former read as~\cite{bonvin:numbercountfluctuation, challinor:numbercountfluctuation, jeong:numbercountfluctuation}
\begin{equation}   
    C^{XY}_\ell(z_i,z_j) = 4\pi \int d\log k \ \mathcal{P}_\mathcal{R}(k) \Delta^{X,z_i}_{\ell}(k) \Delta^{Y,z_j}_{\ell}(k)\, ,
\label{eq:cl} 
\end{equation}
where~$\mathcal{P}_\mathcal{R}$ is the almost scale-invariant primordial curvature power spectrum, and~$k$ and~$\ell$ label Fourier modes and multipoles, respectively.
The harmonic transfer functions are defined as
\begin{equation}
    \Delta^{X,z_i}_{\ell}(k) = \int_0^\infty dz \frac{dN_X}{dz} W(z, z_i, \Delta z_i) \Delta^X_\ell(k,z)\,,
\label{eq:transfers}
\end{equation}
where~$dN_X/dz$ describes the redshift-dependent number distribution of the sources,~$W(z,z_i,\Delta z_i)$ is a window function centered at~$z_i$, with half-width~$\Delta z_i$, and normalized to unity. 

The explicit form of the total harmonic transfer function~$\Delta^X_\ell(k,z)$ contains contributions from density, velocity, and gravitational potential fluctuations.
In particular, the ``density term'' describes how any given class of LSS-probe traces the underlying matter distribution, is the dominant term when~$z_i=z_j$, and it reads as
\begin{equation}
    \Delta^{X,\mathrm{den}}_\ell (k,z) \propto b_X(k,z) D(k,z),
\end{equation}
where~$D(k,z)$ is the gauge-invariant matter fluctuation, and~$b_X(k,z)$ is the linear bias function~\cite{desjacques:biasreview}. 
On the other hand, the angular power spectrum of the noise includes all instrumental, systematic, and intrinsic sources of noise, including the shot noise. 

In this work, we aim to establish whether different clustering models are statistically distinguishable from each other. 
For this purpose, we introduce a Signal-to-Noise ratio defined as 
\begin{equation}
    \mathrm{SNR}^2 = f_{\mathrm{sky}} \sum_{\ell=2}^{\ell_\mathrm{max}} \frac{2\ell+1}{2} \mathrm{Tr} \left[ \Delta\mathcal{C}_\ell \left( \tilde{\mathcal{C}}^\mathrm{fid}_{\ell} \right)^{-1} \Delta\mathcal{C}_\ell \left( \tilde{\mathcal{C}}^\mathrm{fid}_{\ell} \right)^{-1} \right],
\end{equation}
where~$f_\mathrm{sky}$ is the observed fraction of the sky, $\ell_\mathrm{max}$ is the maximum observed multipole, $\mathcal{C}_\ell$ is a covariance matrix containing the angular power spectra, $\Delta\mathcal{C}_\ell = \mathcal{C}^\mathrm{alt}_\ell - \mathcal{C}^\mathrm{fid}_{\ell}$ is the difference between the covariance matrices of the alternative and fiducial time-delay models, while~$\tilde{\mathcal{C}}^\mathrm{fid}_{\ell}$ is the covariance matrix of the fiducial model, which also includes the noise.
The explicit form of the covariance matrix can be found, for instance, in Refs.~\cite{scelfo:gwxlssI, bosi:gwxlssI}.

%%%%%%%%%%%%%%%%%%%%%%%%%%%%%%%%%%%%%%%%%%%%%%%%%%%%%%%%%%%%%%%%%%%%%%%%%%%%%%%%%%%%%%%%%%%%%%%%%%%%%%%%%%%%%%%%%%%%%%%%%%%%%%%%%%%%

\subsection{Selecting the galaxy population}

We expect a cross-correlation signal to exist independently of the chosen galaxy population, since both resolved GWs and galaxies are biased tracers of the dark matter field.
The signal itself approximately scales as~$C^\mathrm{gal,GW}_\ell \approx b_\mathrm{gal} b_\mathrm{GW} C^\mathrm{dm,dm}_\ell$, where~$C^\mathrm{dm,dm}_\ell$ is the dark matter angular power spectrum. Thus, choosing a galaxy sample with large bias certainly helps.
However, an aspect that has received very little attention in this sort of studies is how actually uncorrelated galaxy and GW populations are, i.e., whether shot noise is actually zero for the cross-correlation signal~\cite{alonso:crossshotnoise, cusin:crossshotnoise}.

Let us consider the case in which the observed GWs are sourced from inside the same galaxies targeted by the galaxy survey and used to perform the cross-correlation.
In this case, we expect that
\begin{equation}
    \left\langle \delta_\mathrm{gal} \delta_\mathrm{GW} \right\rangle = \left\langle \left( \frac{N_\mathrm{gal} - \bar{N}_\mathrm{gal}}{\bar{N}_\mathrm{gal}} \right) \left( \frac{N_\mathrm{GW} - \bar{N}_\mathrm{GW}}{\bar{N}_\mathrm{GW}} \right) \right\rangle \neq 0, 
\label{eq:shotnoise_correlatedsample}
\end{equation}
where~$N_\mathrm{gal}$ is the number of galaxies, $N_\mathrm{GW}=\sum_{i=1}^{N_\mathrm{gal}} N_{\mathrm{GW/gal},i}$ is the number of GWs, and~$N_{\mathrm{GW/gal},i}$ is the number of GWs of the~$i$-th galaxy.
The numbers of galaxies and GWs follow a Poissonian and compound Poissonian distribution, respectively, with expectation value~$\left\langle N_\mathrm{gal} \right\rangle = \bar{N}_\mathrm{gal}$ and~$\left\langle N_\mathrm{GW} \right\rangle = \bar{N}_\mathrm{GW} = \bar{N}_\mathrm{gal} \bar{N}_\mathrm{GW/gal}$.
The numerator of equation~\eqref{eq:shotnoise_correlatedsample} reads as
\begin{equation}
    \begin{aligned}
    & \left\langle N_\mathrm{gal} \sum_{i=1}^{N_\mathrm{gal}} N_{\mathrm{GW/gal},i} \right\rangle - \bar{N}_\mathrm{gal} \bar{N}_\mathrm{GW} = \left\langle N^2_\mathrm{gal} \right\rangle \bar{N}_\mathrm{GW/gal} - \bar{N}^2_\mathrm{gal} \bar{N}_\mathrm{GW/gal} = \bar{N}_\mathrm{gal} \bar{N}_\mathrm{GW/gal}; 
    \end{aligned}
\end{equation}
therefore, the cross-shot noise term is given by
\begin{equation}
    \left\langle \delta_\mathrm{gal} \delta_\mathrm{GW} \right\rangle = \frac{1}{\bar{N}_\mathrm{gal}}.
\end{equation}
In contrast, assuming that galaxies can be distinguished into two types, A and B, and that only the latter type hosts GW events, the expectation value  of galaxy of type A and GWs coming from galaxies of type B is
\begin{equation}
    \left\langle \delta^A_\mathrm{gal} \delta^B_\mathrm{GW} \right\rangle = \left\langle \left( \frac{N^A_\mathrm{gal} - \bar{N}^A_\mathrm{gal}}{\bar{N}^A_\mathrm{gal}} \right) \left( \frac{N^B_\mathrm{GW} - \bar{N}^B_\mathrm{GW}}{\bar{N}^B_\mathrm{GW}} \right) \right\rangle = 0,
\end{equation}
since we can factorize the two ensemble averages over the numbers of galaxies and GWs.
The same logic suggests that also in the case of cross-correlating galaxies and GWs in non-overlapping redshift bins we expect zero cross-shot noise, similarly to what happens in standard galaxy-galaxy and GW-GW scenario.
With partially overlapping redshift bins, we are in an intermediate situation, where the magnitude of the cross-shot noise term is reduced with respect to the perfectly overlapping bins case.

In a more realistic scenario, where the GW number count fluctuation is constructed from events coming from both galaxy populations, we have that
\begin{equation}
    \delta_\mathrm{GW} = \frac{N^A_\mathrm{GW}}{N^\mathrm{tot}_\mathrm{GW}} \delta^A_\mathrm{GW} + \frac{N^B_\mathrm{GW}}{N^\mathrm{tot}_\mathrm{GW}} \delta^B_\mathrm{GW},
\end{equation}
where~$N^A_\mathrm{GW} + N^B_\mathrm{GW} = N^\mathrm{tot}_\mathrm{GW}$; thus, when computing the expected level of cross-shot noise, we find
\begin{equation}
    \left\langle \delta_\mathrm{GW} \delta^A_\mathrm{gal} \right\rangle = \frac{N^A_\mathrm{GW}}{N^\mathrm{tot}_\mathrm{GW}} \frac{1}{N^A_\mathrm{gal}} \leq \frac{1}{N^A_\mathrm{gal}}.
\label{eq:crosscorrelation_shotnoise}
\end{equation}
We conclude that, from a purely theoretical standpoint, in order to minimize the noise term appearing in equation~\eqref{eq:cl_general} and to maximize the information content contained in the clustering signal, it is highly preferable to cross-correlate GWs with a galaxy population that does not host GWs.

%%%%%%%%%%%%%%%%%%%%%%%%%%%%%%%%%%%%%%%%%%%%%%%%%%%%%%%%%%%%%%%%%%%%%%%%%%%%%%%%%%%%%%%%%%%%%%%%%%%%%%%%%%%%%%%%%%%%%%%%%%%%%%%%%%%%

\subsection{The role of delay time}
\label{subsec:timedelay}

The reasoning in the previous section suggests a peculiar interplay between the typical GW time-delay and the typical duration of the star-forming phase of a galaxy.
In what follows, we distinguish between two types of galaxies: red/quiescent (QG), and blue/star-forming (SFG).
Additionally, for the purpose of estimating the expected level of cross-shot noise, we reason in terms of the average GW time-delay~$\bar{t}_d$ and galaxy active-star forming time~$\bar{t}_\mathrm{SF}$.
Other estimators, such as median times, can also be adopted, especially if the distributions of these times have wide tails; however, we leave the detailed exploration of alternative estimators for future work.

For the sake of providing a qualitative argument, let us consider a time-delay probability distribution function (pdf) parametrized as~$p(t_d)\propto t_d^{\alpha_d}$.
In this scenario, the average time-delay is
\begin{equation}
    \bar{t}_d = 
    \left\lbrace \begin{aligned}
    & \frac{\log(t_{d,\mathrm{max}}/t_{d,\mathrm{min}})}{t^{-1}_{d,\mathrm{min}} - t^{-1}_{d,\mathrm{max}}}, \quad & \alpha_d = -2, \\
    & \frac{t_{d,\mathrm{max}} - t_{d,\mathrm{min}}}{\log(t_{d,\mathrm{max}}/t_{d,\mathrm{min}})}, \quad & \alpha_d = -1, \\
    & \frac{\alpha_d+1}{\alpha_d+2} \frac{ t^{\alpha_d+2}_{d,\mathrm{max}} - t^{\alpha_d+2}_{d,\mathrm{min}} }{ t^{\alpha_d+1}_{d,\mathrm{max}} - t^{\alpha_d+1}_{d,\mathrm{min}} }, \quad & \alpha_d \neq -2, -1, \\
    \end{aligned} \right.
\end{equation}
where~$t_{d,\mathrm{min}}, t_{d,\mathrm{max}}$ are the minimum and maximum time-delay, respectively.
If~$\bar{t}_d \lesssim \bar{t}_\mathrm{SF}$, it is very likely that we observe a GW coming from a galaxy that is still in its star-forming era.
Conversely, if~$\bar{t}_d \gtrsim \bar{t}_\mathrm{SF}$, it is very likely that observed GWs come from QGs.
Therefore, if we assume that~$\bar{t}_\mathrm{SF} \approx \mathrm{few\ Gyr}$~\cite{lian:galaxyphases}, $t_{d,\mathrm{min}} \approx 2\ \mathrm{Myr}$, and~$t_{d,\mathrm{max}}=14\ \mathrm{Gyr}$, a quick estimate tells us that if~$\alpha_d \leq -1$ ($\alpha_d \geq -1$), it is desirable to use a sample of red (blue) galaxies to minimize the cross-correlation shot noise.
Finally, since at low redshift the transition between the star-forming and quiescent stages occurs very rapidly, this phenomenological separation has the potential to be quite sharp~\cite{lian:galaxyphases}. 
The situation may change at high redshift, where recent JWST observations suggest that increased burstiness may drive variations in galaxy star formation histories and in the time-scales over which star formation and feedback operate; see, e.g., the discussion in Ref.~\cite{Sun:2023ocn} and references therein.
Since these redshifts are beyond the reach of current GW interferometers and will be probed only by a small fraction of GW events by next-generation detectors, we leave a detailed treatment of this possibility for future work.

\begin{table}[ht]
    \centerline{
    \begin{tabular}{|c|c|c|c|c|c|}
    \hline
    $\alpha_d$ & -2.0 & -1.5 & -1.0 & -0.5 & 0.0 \\
    \hline
    \hline
    $\mathrm{CDF}(\bar{t}_\mathrm{SF})$ & $\gtrsim 0.99$ & $0.98$ & $0.80$ & $0.45$ & $0.22$ \\
    \hline
    \end{tabular}}
    \caption{Time-delay cumulative distribution function for~$\bar{t}_\mathrm{SF}=3\ \mathrm{Gyr}$, corresponding to the fraction of GW events observed in galaxies still in their star-forming stage.}
    \label{tab:timedelay_cdf}
\end{table}

In the following, we explore how the GW$\times$LSS cross-correlation changes for different GW time-delay distributions, when the galaxy survey targets either a SFG or a QG population.
Since we want to understand the impact of the cross-correlation shot noise introduced in equation~\eqref{eq:crosscorrelation_shotnoise}, we consider the same properties for both populations, i.e., same total number of objects, redshift distribution, and bias functions, which we report in appendix~\ref{app:tracer_population}.
To compute the cross-shot noise in the case where the galaxy survey targets an SFG population, we need to estimate the probability that a GW event, which is very likely sourced in an SFG, is observed in a galaxy still in its star-forming phase.
Otherwise said, we are interested in the average ratio between the number of GW events observed in an SFG with respect to the total number of GWs, i.e.,~$N^\mathrm{SFG}_\mathrm{GW}/N^\mathrm{tot}_\mathrm{GW}$.
This fraction is simply given by the time-delay cumulative distribution function~$\mathrm{CDF}(\bar{t}_\mathrm{SF})$, which we report in table~\ref{tab:timedelay_cdf}.
On the other hand, in the case where we cross-correlate GW with a QG population, the number of GW born in SFGs and observed in QGs is simply given by~$1-\mathrm{CDF}(\bar{t}_\mathrm{SF})$.

%%%%%%%%%%%%%%%%%%%%%%%%%%%%%%%%%%%%%%%%%%%%%%%%%%%%%%%%%%%%%%%%%%%%%%%%%%%%%%%%%%%%%%%%%%%%%%%%%%%%%%%%%%%%%%%%%%%%%%%%%%%%%%%%%%%%

\section{Constraining binary formation channels}
\label{sec:binary_formation_channels}

Recent detections of GW events point towards a picture where BBHs originate from multiple formation pathways.
In addition to mass and spin distributions, the merger rate density could also help shed light on the presence of multiple BBH populations. 
However, a robust theoretical model linking the observed merger rate to the underlying properties of GW sources and their host galaxies is needed to take advantage of this probe.
Among host properties, the stellar mass ($M_*$) is the most prominent~\cite{Artale:2019doq, Artale:2019tfl}, although star formation (SFR) and metallicity ($Z$) have also been shown to play an important role~\citep{Langer:2005hu, Ma:2015nya, Chruslinska:2018hrb, boco:metallicitypdf}.
The BBH formation rate is commonly set by a metallicity-dependent star formation history kernel, which can be derived through cosmological simulations, see, e.g., Refs.~\cite{Mapelli:2017hqk, Schneider:2017rdg, lamberts:mr, artale:mr, levina:mr}, or by observation-driven parametric relations among the host properties, such as stellar mass, SFR, and metallicity, see, e.g., Refs.~\cite{dominik:mr, belczynski:mr, neijssel:mr, santoliquido:host, broekgaarden:mr, sgalletta:sfrsb, boco:mr}.
The mapping from formation to coalescence also employs a time-delay pdf, which describes the time required for a given binary to merge after its formation.

Providing an accurate model for the time-delay pdf represents a challenging endeavor.
As the total cosmic star formation density embeds contributions from distinct environments and BBH formation channels, the shape of the average time-delay pdf is determined by the relative importance of each individual BBH formation pathway and host environment~\cite{mapelli:fastclusterII,sedda:bpop}.
The latter influences~$p(t_d)$ mainly through the metallicity, since short (long) time-delays have been shown to be associated with environments with low (high) metallicity~\cite{marchant:ch, duBuisson:2020asn, guerrero:tdmet}.
Regarding the formation channel, the time-delay pdf of isolated binaries that evolve through mass transfer usually follows a power-law distribution, i.e.,~$p(t_d)\propto t_d^{\alpha_d}$, with a typical value of~$\alpha_d = -1$ derived from the assumption that the BBH semi-major axis at formation, $a$, follows a distribution~$p(a) \propto a^{-1}$~\cite{Abt:1983tr, sana:tdiso}.

However, the domain of validity of this assumption is highly uncertain and likely depends on the treatment of metallicity~\cite{lamberts:met}, stellar winds, mass transfer, or natal kicks~\cite{OShaughnessy:2007brt, OShaughnessy:2009szr, Mapelli:2017hqk, fishbach:tdiso}.
Different time-delay distributions are predicted also in the scenario of isolated-BBH formation: even assuming a chemically homogeneous evolution and a stable mass transfer, the triple system channel might lead to longer time-delays, depending on metallicity and natal kicks~\cite{mandel:chetd, Marchant:2016wow, duBuisson:2020asn, deMink:chetd, antonini:triples}.
On the other hand, dynamical interactions in dense environments influence binary evolution: mergers occurring within globular clusters tend to experience short time-delays, which are, however, significantly increased if the BBH is ejected from the cluster before merging~\citep{Benacquista:2011kv, Rodriguez:2016kxx, Banerjee:2016ths, Rodriguez:2017pec}.
Short time-delays have also  been shown to be associated with BBHs formed inside young star clusters~\citep{DiCarlo:2020lfa} or in AGN disks~\citep{Yang:2020lhq}, where the time-delay pdf shape and tilt may appreciably vary depending on which process drives the merger~\cite{vaccaro:agn}.

Given these large uncertainties in the different theoretical models for BBH formation, as well as the unknown relative importance of each mechanism, it appears natural to look for alternative probes able to test these hypotheses. 
In this section, we explore the potential of GW$\times$LSS cross-correlation to measure the response of the GW bias to different time-delay pdfs, as well as the combined effect of BBHs forming in different environments with different time-delay distributions.
Firstly, we consider a model-agnostic framework where the time-delay distribution is given by the parametric form~$p(t_d)\propto t_d^{\alpha_d}$ already considered in section~\ref{subsec:timedelay}, without linking it to any specific formation mechanism.
Secondly, we employ a more detailed multi-population scenario where the time-delay pdf is derived from population synthesis codes and based on a first-principle approach to stellar evolution.

%%%%%%%%%%%%%%%%%%%%%%%%%%%%%%%%%%%%%%%%%%%%%%%%%%%%%%%%%%%%%%%%%%%%%%%%%%%%%%%%%%%%%%%%%%%%%%%%%%%%%%%%%%%%%%%%%%%%%%%%%%%%%%%%%%%%

\subsection{Model-agnostic scenario}
\label{subsec:model_agnostic_estimate}

The GW merger rate is set by the convolution between the time-delay distribution and the binary formation rate, see, e.g., equation~\eqref{eq:merger_rate} in appendix~\ref{app:BBH_formation}. 
In the presence of multiple formation pathways, the observed rate therefore depends on their relative weights, which remain poorly constrained by current models.
Since neither of the two terms appearing in the convolution can be independently probed by GW observation alone, increasing the precision of GW measurements cannot overcome this intrinsic degeneracy.
Therefore, breaking such a degeneracy requires one to measure a new observable that, acting as a ``second clock'' along with the merger rate, allows us to disentangle the effect of the time-delay pdf from the binary formation rate.
The GW bias represents one viable option that will be available in the near future.

\begin{figure}[ht]
    \centerline{
    \includegraphics[width=\columnwidth]{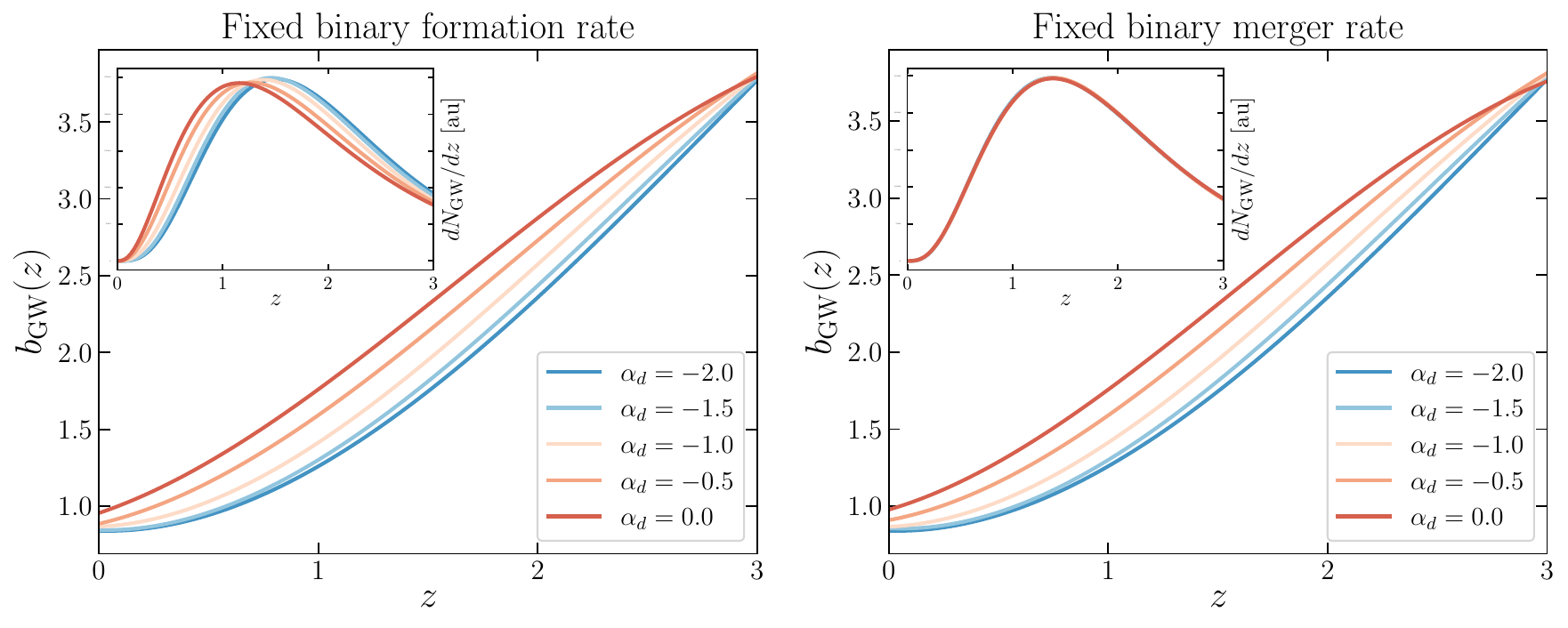}}
    \caption{\textit{Small panels:} GW number density distribution in arbitrary units [au] for different choices of the time-delay power-law distribution, and fixing either the BBH formation rate (\textit{left panel}) or their merger rate (\textit{right panel}). 
    \textit{Large panels:} GW bias for the same two scenarios and choice of time-delay pdf exponents.}
    \label{fig:mergerrate_degeneracy}
\end{figure}

To showcase the potential of GW clustering in breaking this degeneracy, we consider two different scenarios, a ``fixed binary formation rate'' and a ``fixed merger rate'' one.
In both scenarios, the time-delay pdf is parametrized as~$p(t_d) \propto t_d^{\alpha_d}$, where we explore the range of exponents~$\alpha_d=\{-2.0,-1.5,-1.0,-0.5,0.0\}$ and consider~$\alpha_d=-1$ as our fiducial model for the purpose of inferring the distinguishability between models.
Small (large) exponents correspond to shorter (longer) coalescence times.
Consequently, in the scenario in which we fix the binary formation rate to its fiducial value, different exponents are responsible for different observed merger rates.
Similarly, in the fixed merger rate scenario, different exponents of the time-delay pdf will be associated with different binary formation rates to provide the same merger rate. 
In both scenarios, we fix the local merger rate to be consistent with the observed LIGO-Virgo-KAGRA value.

Following Ref.~\cite{bellomo:classgwb}, we generate catalogs of GW events detected by a third-generation detector network for each scenario and value of the exponent; in particular, we consider the combination of Einstein Telescope and two Cosmic Explorers, as explained in appendix C of Ref.~\cite{bosi:gwxlssI}.
In each catalog, we assign a GW event to a specific halo via a pdf based on the SFR-halo mass relation provided by the \textsc{UniverseMachine} framework~\cite{behroozi:universemachine}, see also Ref.~\cite{bellomo:classgwb} for a more detailed step-by-step explanation.
The GW bias is then calculated using an HOD model~\cite{libanore:gwxlssI, bellomo:classgwb}.
We show both the GW number densities and the biases in figure~\ref{fig:mergerrate_degeneracy}.
Most notably, we observe that also in the fixed merger rate scenario, where the redshift distribution is the same for all time-delay pdfs, the GW bias is not degenerate, i.e., we can observe the effect of different time-delay pdfs in this second clock.

\begin{figure}[ht]
    \centerline{
    \includegraphics[width=\columnwidth]{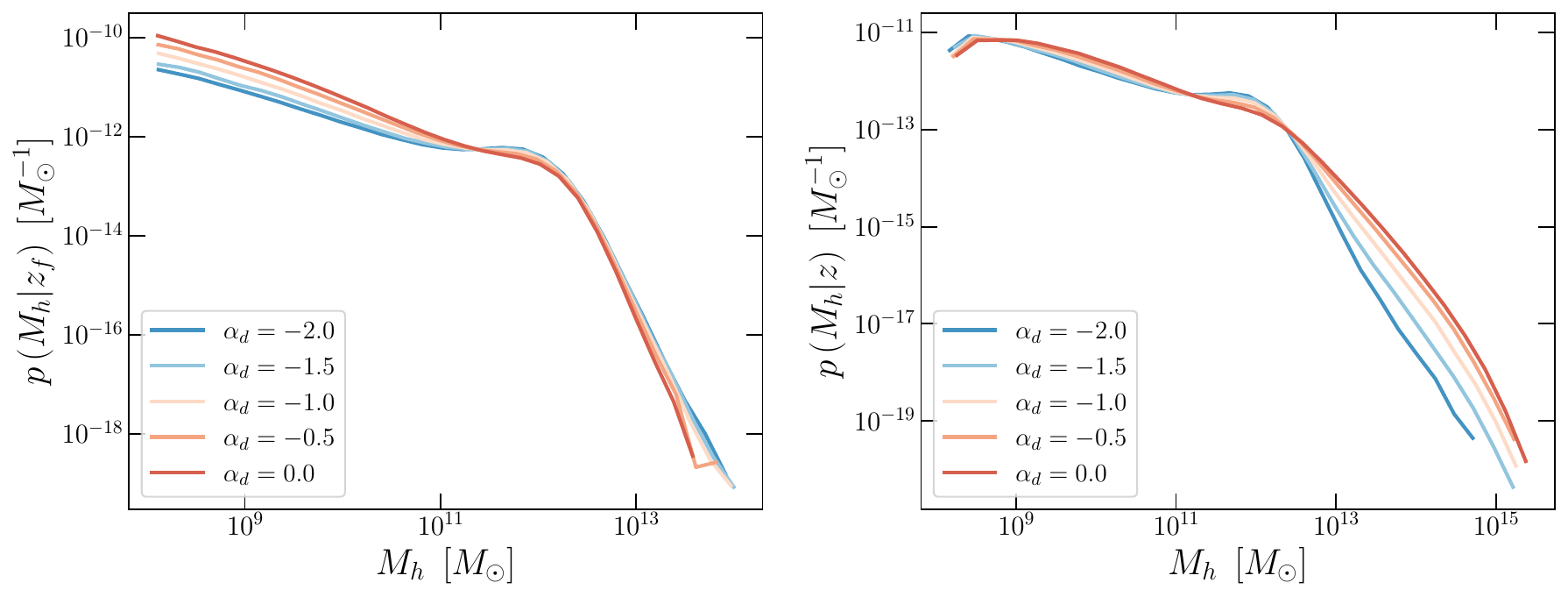}}
    \caption{Probability distribution function of the mass of the halo hosting a GW events at formation (\textit{left panel}) and merger (\textit{right panel}) for the fixed merger rate scenario and different time-delay distribution.
    As expected, the larger the exponent, the larger the mass of the host at the merger redshift; thus, the larger the bias.}
\label{fig:p_Mh}
\end{figure}

The explanation of this effect can be immediately appreciated by looking at figure~\ref{fig:p_Mh}, where we report the pdf of the masses of the halos hosting a GW event at binary formation redshift~$z_f$ and binary merger redshift~$z$.
Since the GW bias is a weighted average of the halo bias, where the weight is given by the probability of observing a GW event in a given halo reported in the right panel of the figure, these pdfs naturally indicate which type of host dominates the average.
In other words, the GW bias acts as an indicator of which dark matter halo is most likely to host a merging BBH, which in turn depends on how GWs trace the underlying LSS.
Therefore, different time-delay distributions naturally select a different population of hosting halos at binary formation redshift, as we observe in the left panel of the figure.
Additionally, regardless of where the BBHs originally form, longer time-delays allow dark matter halos to grow more before the merger takes place; thus, larger exponents favor longer time-delays and result in more massive host at merger, i.e., in higher GW bias.

\begin{figure}[ht]
    \centerline{
    \includegraphics[width=\columnwidth]{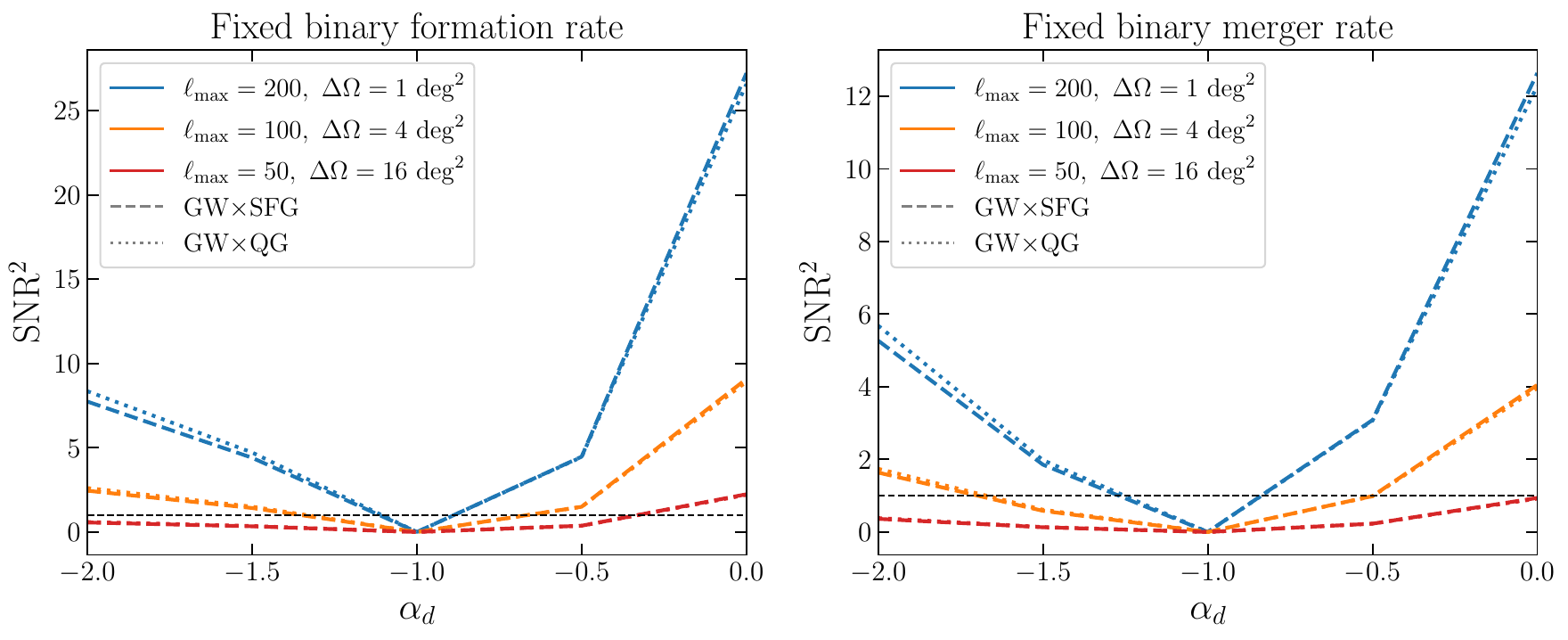}}
    \caption{SNR$^2$ for different time-delay distributions~$p(t_d)\propto t_d^{\alpha{_d}}$ with respect to the fiducial case~$\alpha_d=-1$, for the fixed binary formation rate (\textit{left panel}) and fixed merger rate (\textit{right panel}) scenarios.
    \textit{Blue}, \textit{orange}, and \textit{red lines} represent scenarios with different average uncertainties on the GW localization.
    \textit{Dashed} and \textit{dotted lines} represent cross-correlations performed with SFG and QG populations, respectively.}
\label{fig:agnostic_ptd}
\end{figure}

Once established that the information about the time-delay pdf is intrinsically encoded in the clustering properties of the GW samples, we turn our attention to the possibility of discriminating between different models.
In particular, we consider the impact of cross-correlating GWs with both SFG/QG populations and the impact of limited angular resolution for GW events. 
We bin the GW population as explained in appendix~\ref{app:tracer_population}, and we compute the angular power spectra of GWs, galaxies and their cross-correlation using the most recent release of~\texttt{Multi\_CLASS}~\cite{bellomo:forecastapproximations, bernal:multiclass, scarpel:gwxlss}.
Regarding the spatial resolution of GW events, we assume an average angular uncertainty of~$\Delta\Omega=1,4, 16\ \mathrm{deg}^2$, thus restricting our angular analysis to a maximum multipole of~$\ell_\mathrm{max}=200,100,50$.
We report in figure~\ref{fig:agnostic_ptd} the statistical distinguishability for both the fixed binary formation rate and the fixed merger rate scenarios, for different values of the time-delay pdf exponent.
Two models are considered statistically distinguishable when~$\mathrm{SNR}^2\geq 1$.

First, we observe that, as expected, SFGs (QGs) are more effective in constraining scenarios with longer (shorter) delay times.
The magnitude of this effect in this case does not appear to be significant, since most of the constraining power is still retained in the cross-tracer cross-bin angular power spectra.
However, for different survey configurations with a smaller number of cross-tracer cross-bin measurements, the cross-shot noise can easily have a larger impact. 
Fixed binary formation scenarios typically have larger~$\mathrm{SNR}^2$: also this effect is expected, since these models have different GW redshift distributions, which act as different weights of the bias functions, thus enhancing the inter-model differences.
Regardless of this aspect, we observe that in the fixed merger rate scenario we maintain the distinguishability between models, especially when the average resolution of the events drops below~$\Delta\Omega \lesssim 5\ \mathrm{deg}^2$.
In other words, our forecasts show that, across most of the astrophysical models considered, the imprint of the time-delay distribution on the GW clustering is robustly detectable, even without adding additional information regarding the specific type of structure each model is connected to.
Therefore, this result represents the first hint about how the cross-correlation between GWs and galaxies can independently probe the physics of BBH formation. 

%%%%%%%%%%%%%%%%%%%%%%%%%%%%%%%%%%%%%%%%%%%%%%%%%%%%%%%%%%%%%%%%%%%%%%%%%%%%%%%%%%%%%%%%%%%%%%%%%%%%%%%%%%%%%%%%%%%%%%%%%%%%%%%%%%%%

\subsection{Multi-population scenario}
\label{subsec:multi_population_scenario}

Let us now consider a scenario where GWs are explicitly sourced by two different BBH formation channels, i.e., isolated binary evolution (IB) and globular cluster formation (GC), each one characterized by a unique environment and time-delay pdf, and analyze whether we can disentangle their presence via a GW clustering measurement.
In this case, we adopt a first principle approach based on numerical simulations of the evolution of BBH populations.
Specifically, we use~\textsc{SEVN}~\cite{spera:sevn, iorio:sevn} and~\textsc{FastCluster}~\cite{mapelli:fastclusterI, mapelli:fastclusterII, torniamenti:fastcluster} to create BBH catalogs for the IB and GC scenarios, respectively. 
For each channel, we run~$15$ simulations at different metallicity, from~$Z_\mathrm{sim}=0.0002$ to~$Z_\mathrm{sim}=0.02$.
This formalism is general enough to account for BBH formed in other kinds of stellar cluster environment, which also may play a role; however, for simplicity, we limit our proof-of-concept analysis to GCs.

First, we build a synthetic realization of our Universe as detailed in appendix~\ref{app:synthetic_universe}.
Our formalism is able to self-consistently simulate a population of dark matter halos, and to assign them a galaxy population in such a way that the observed galaxy stellar mass function, star-formation rate, metallicity, and GC distributions are consistent with observations.
Thanks to BBH simulations and our semi-analytical approach, we construct for each formation channel~$\{ \mathrm{ch} \} = \{ \mathrm{IB}, \mathrm{GC} \}$ the intrinsic merger rate density
\begin{equation}
    R^\mathrm{ch}_\mathrm{M}(z) = \int^{t(z)}_0 dt_d dZ_f p_\mathrm{ch}(t_d|Z_f) \varepsilon_\mathrm{ch}(Z_f) \frac{\dot{\rho}^\mathrm{ch}_\mathrm{SFR} (Z_f, z_f)}{\left\langle m_\star \right\rangle} ,
\end{equation}
where~$z$ is the merger redshift, $t(z)$ is the cosmic time at redshift~$z$, $z_f$ is the binary formation redshift, $Z_f$ is the metallicity of the environment where the binary forms, and $\left\langle m_\star \right\rangle$ is the average stellar mass. For each channel (ch) we define $p_{\rm ch}(t_d|Z_f)$ as the time-delay pdf at a given metallicity, $\varepsilon_{\rm ch}$ as the binary formation efficiency, and~$\dot{\rho}^\mathrm{ch}_\mathrm{SFR}$ as the star-formation rate of the channel.
The values of~$\varepsilon_\mathrm{ch}$ and~$p_\mathrm{ch}$ are obtained from the simulations, as detailed in appendix~\ref{app:BBH_formation}.
In particular, we find that these two formation channels are responsible for quite different time-delay distributions, see, e.g., figures~\ref{fig:IB_ptd_efficiency} and~\ref{fig:GC_ptd_efficiency} of appendix~\ref{app:BBH_formation}, making them the perfect candidate to test the potential of GW clustering.

Regarding the total star-formation rate, we assume it is given by two contributions, i.e.,
\begin{equation}
    \dot{\rho}^\mathrm{tot}_\mathrm{SFR}(Z_f,z_f) = \dot{\rho}^\mathrm{IB}_\mathrm{SFR}(Z_f,z_f) + \dot{\rho}^\mathrm{GC}_\mathrm{SFR}(Z_f,z_f),
\end{equation}
where~$\dot{\rho}^\mathrm{GC}_\mathrm{SFR}(Z_f,z_f) = f_\mathrm{GC}(z_f) \dot{\rho}^\mathrm{tot}_\mathrm{SFR}(Z_f,z_f)$, $f_\mathrm{GC}(z_f) = \dot{\rho}^\mathrm{GC}_\mathrm{SFR} / \dot{\rho}^\mathrm{tot}_\mathrm{CSFR}$, the GC star-formation rate is given in Ref.~\cite{elbadry:globularclustersimulation}, and the total cosmic star-formation rate is matched to observations.
Therefore, the IB channel star-formation rate density is given by construction as~$\dot{\rho}^\mathrm{IB}_\mathrm{SFR}(Z_f,z_f) = \left[ 1 - f_\mathrm{GC}(z_f) \right] \dot{\rho}^\mathrm{tot}_\mathrm{SFR}(Z_f,z_f)$, as done in Ref.~\cite{bosi:mergereff}.
Additionally, we also adjust the overall values of the efficiency of each channel to match the observed value of the local merger rate, while keeping the shape predicted by numerical simulations.
Finally, we construct a comprehensive distribution of GW events coming from both formation channels as
\begin{equation}
    \frac{dN^\mathrm{tot}_\mathrm{GW}}{dz} = f_\mathrm{IB} \frac{dN^\mathrm{IB}_\mathrm{GW}}{dz} + \left( 1 - f_\mathrm{IB} \right) \frac{dN^\mathrm{GC}_\mathrm{GW}}{dz}
\end{equation}
where~$f_\mathrm{IB}\in [0,1]$ indicates the fraction of events coming from the IB channel.

Since the BBH formation processes occur in different environments, galaxies as a whole versus GC specifically, it seems very likely that GW clustering might receive a clear imprint of such connection.
As showed in appendix~\ref{app:synthetic_universe}, galaxies and GCs trace slightly differently the underlying dark matter distribution, i.e., they have different biases, see, e.g., the left panel of figure~\ref{fig:GSMF_bias}.
Thanks to the halo-occupation distribution (HOD) formalism introduced in the appendix~\ref{app:synthetic_universe}, we can easily compute the GW bias of each individual channel as
\begin{equation}
    \begin{aligned}
        b^\mathrm{IB}_\mathrm{GW} &= n^{-1}_\mathrm{GW} \int dM_h dM_\star \frac{dn_h}{dM_h} \left\langle \frac{dN_\mathrm{gal}}{dM_\star} \bigg| M_h \right\rangle \left\langle N_\mathrm{GW/gal} | M_h,M_\star \right\rangle b_h(M_h), \\
        b^\mathrm{GC}_\mathrm{GW} &= n^{-1}_\mathrm{GW} \int dM_h dM_\mathrm{GC} \frac{dn_h}{dM_h} \left\langle \frac{dN_\mathrm{GC}}{dM_\mathrm{GC}} \bigg| M_h \right\rangle \left\langle N_\mathrm{GW/GC} | M_h,M_\mathrm{GC} \right\rangle b_h(M_h),
    \end{aligned}
\label{eq:formation_mechanism_bias}
\end{equation}
where~$n_\mathrm{GW}$ is the GW number density, $dn_h/dM_h$ is the halo mass function, $dN_\mathrm{gal}/dM_\star$ is the number of galaxies per stellar mass bin, $dN_\mathrm{GC}/dM_\mathrm{GC}$ is the number of GC per GC mass bin, $N_\mathrm{GW/gal}$ is the number of GW per galaxy, $N_\mathrm{GW/GC}$ is the number of GW per galaxy, and~$b_h$ is the halo bias.
In principle, the HOD model can be further extended by including dependencies on the star-formation rate~\cite{libanore:gwxlssI} and the metallicity~\cite{boco:biasmodel,scelfo:gwxlssII}. 
Ref.~\cite{peron:gwxlss} demonstrated that the analytical HOD bias calibrated on the simulated BBH catalogs well matches the numerical estimate of the bias extracted from the simulation themselves. 
As for this work, dependencies on these additional parameters are intrinsically present, since they are naturally integrated in the process of realizing a synthetic Universe that matches observations; however, to keep our formalism as clean as possible, we do not make them explicit in the above equations.

\begin{figure}[ht]
    \centerline{
    \includegraphics[width=\columnwidth]{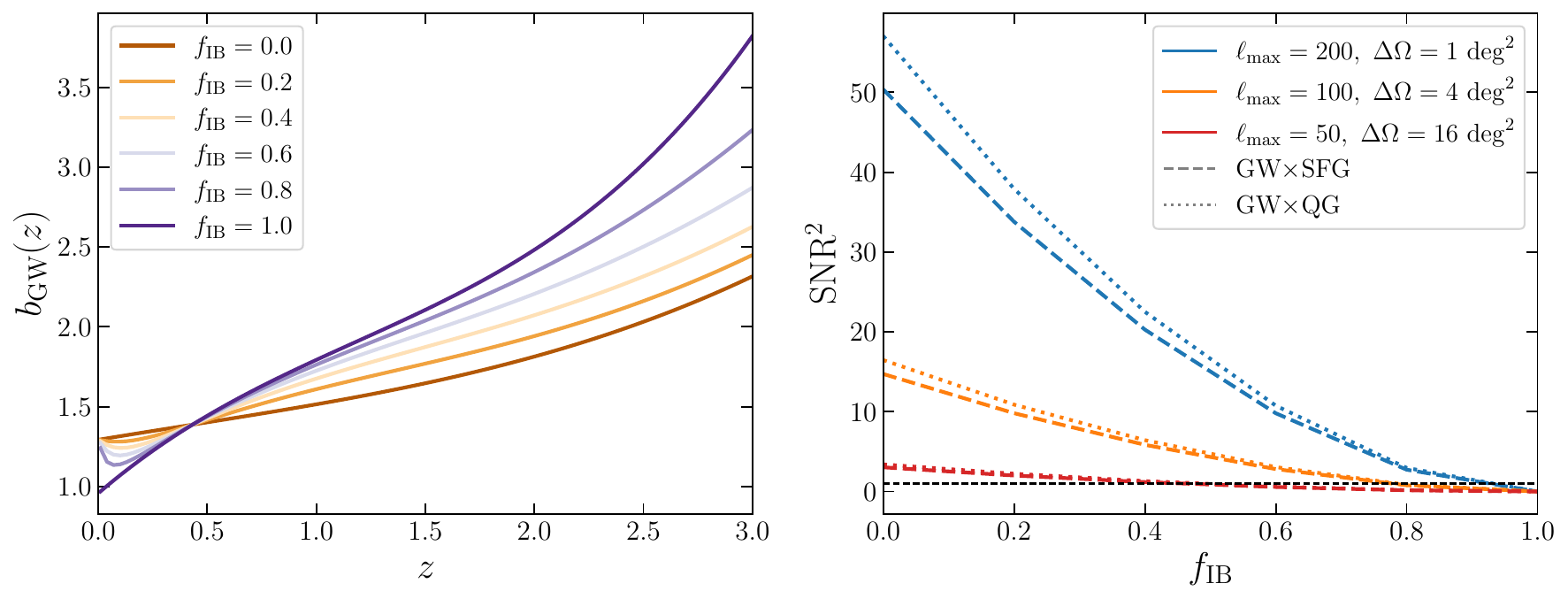}}
    \caption{\textit{Left panel}: GW bias for different mixture of IB/GC channels.
    The case with~$f_\mathrm{IB}=0$ ($f_\mathrm{IB}=1$) corresponds to the case where only the GC (IB) channel contributes.
    \textit{Right panel}: SNR$^2$ for different mixture of IB/GC channels with respect to the fiducial case~$f_\mathrm{IB}=0$.
    \textit{Blue}, \textit{orange}, and \textit{red} lines represent scenarios with different average uncertainty on the GW localization.
    \textit{Dashed} and \textit{dotted lines} represent cross-correlations performed with SFG and QG populations, respectively.}
\label{fig:SNR2_IB_GC}
\end{figure}

The total bias of the GW population is also given by the sum of two contributions, i.e., 
\begin{equation}
    b^\mathrm{tot}_\mathrm{GW} = \left( \frac{dN^\mathrm{tot}_\mathrm{GW}}{dz} \right)^{-1} \left[ f_\mathrm{IB} \frac{dN^\mathrm{IB}_\mathrm{GW}}{dz} b^\mathrm{IB}_\mathrm{GW} + \left( 1 - f_\mathrm{IB} \right) \frac{dN^\mathrm{GC}_\mathrm{GW}}{dz} b^\mathrm{GC}_\mathrm{GW} \right].
\end{equation}
We show the GW bias for the total population in the left panel of figure~\ref{fig:SNR2_IB_GC}.
As we observe from the figure, the environment clearly leaves an imprint on the shape and magnitude of the bias function. 
In other words, since the properties of the local environment that regulate BBHs formation and merger are different in the two scenarios, each channel acts preferentially in different types of dark matter halos.
In statistical terms, the halo mass in the IB scenario is sampled from the conditional probability
\begin{equation}
    M_h \sim p_\mathrm{IB}(M_h | M_\star, \mathrm{SFR}, Z_f) \propto \left. \frac{d^4n_\mathrm{gal}}{dM_h dM_\star d\mathrm{SFR} dZ_f} \right|_{M_\star, \mathrm{SFR}, Z_f},
\end{equation}
whereas, in the GC scenario, is sampled from
\begin{equation}
    M_h \sim p_\mathrm{GC}(M_h | M_\mathrm{GC}, Z_f) \propto \left. \frac{d^3n_\mathrm{GC}}{dM_h dM_\mathrm{GC} dZ_f} \right|_{M_\mathrm{GC}, Z_f}.
\end{equation}
Thus, according to our HOD, each channel will naturally have a slightly different bias.

As before, we perform a model distinguishability test to investigate whether these bias differences are detectable or not.
GW events are still binned as explained in appendix~\ref{app:tracer_population}, and angular power spectra are still computed using the most recent release of~\texttt{Multi\_CLASS}.
The result of our test is reported on the right panel of figure~\ref{fig:SNR2_IB_GC}, also in this case assuming that the total number of GW events is~$N_\mathrm{GW} \approx 380\ 000$.
Following the previous logic, we consider both SFG and QG populations with different average angular sensitivities.
We observe a rather promising constraining power in constraining even a fraction of order few ten percent of GWs coming from GCs, especially as the angular resolution decreases below~$\Delta\Omega \lesssim 5\ \mathrm{deg}^2$.
In this case, QGs perform slightly better than SFGs since the metallicity-averaged time-delay cumulative distribution function is~$\mathrm{CDF}_\mathrm{IB}(\bar{t}_\mathrm{SF}) \approx 0.7$ and~$\mathrm{CDF}_\mathrm{GC}(\bar{t}_\mathrm{SF}) \approx 0.95$ for the IB and GC channels, respectively.

%%%%%%%%%%%%%%%%%%%%%%%%%%%%%%%%%%%%%%%%%%%%%%%%%%%%%%%%%%%%%%%%%%%%%%%%%%%%%%%%%%%%%%%%%%%%%%%%%%%%%%%%%%%%%%%%%%%%%%%%%%%%%%%%%%%%

\section{Future outlook and conclusions}
\label{sec:outlook_conclusions}

Despite being a young field, GW astronomy is rapidly improving its ability to trace the LSS of the Universe and promises to become a powerful cosmological probe. 
However, as the precision of the observations increases, addressing the uncertainties of theoretical models and accounting for the variety of scenarios that lead to the formation and merger of BBHs will become the name of the game. 
Given the multiplicity of possible alternative explanations, the mere fit of the data without a careful analysis of the underlying assumptions will severely limit the interpretability. 
Ultimately, it will become crucial and necessary to use multiple observables to distinguish between competing scenarios.

In this paper, we explored how tools inherited from Cosmology can address a fundamentally astrophysical problem. 
In particular, we focused on the clustering of GW events, so far mainly used to tighten the constraining power on cosmological parameters or to identify signatures of the existence of primordial black holes.
Here, instead, we use GW clustering to probe different astrophysical formation channels, thanks to its sensitivity to the BBH time-delay pdf, which in turn is strongly influenced by the binary formation mechanism, and the unique properties of the environment where binaries form.

We relied on simulated state-of-the-art catalogs to show that, regardless of the binary formation channel, models that favor longer time-delays lead to larger GW clustering bias. 
This is a consequence of the larger mass of the underlying dark matter halo hosts, whose mass significantly increases over a shorter timescale compared to the merger time in such scenarios. 
We then specialized our analysis with respect to binaries that form and merge either isolated in the field, or that dynamically bound in high density environments, such as globular clusters. 
Also in this scenario, we found that future GW observatories and galaxy surveys have enough constraining power to discriminate the presence of multiple formation channels.

We also commented on the best strategy to choose the galaxy survey to cross-correlate GW datasets with: if the goal is to constrain the clustering properties, both the galaxy shot noise and the cross-shot noise have to be minimized. 
To do so, the ideal galaxy population is the one that hosts the minimum number of GW events and, possibly, has a large bias to increase the magnitude of the cross-correlation signal.
Since the presence of BBH mergers in short time-delay scenarios is favored in star forming environment, a compelling candidate for the cross correlation are red, quiescent galaxies. 
Viceversa, the best candidate for models with long time-delays is a population of blue, star-forming galaxies.
In this respect, we note that future galaxy surveys might target multiple populations at the same time, providing both options at the same time, as in the case of SKAO, which will detect both a population of radio-loud AGNs and star-forming galaxies.

The formalism developed in this paper can be further generalized to account for additional properties characterizing the BBH population, for instance by creating subcatalogs of GW with different progenitor BH masses. 
Current models indicate that, while isolated black holes can hardly have masses larger than~$40\,\mathrm{M}_\odot$, inside globular clusters the multiple encounters and mergers can lead to the formation of black holes that populate the entire mass spectrum, including the large-mass tail of the BH mass distribution. 
In other words, by comparing the GW bias of maps obtained from different subcatalogs of events, we can either perform a self-consistency check or develop a test to infer the presence of multiple formation channels.
This is particularly in light of the recent findings of Ref.~\cite{bosi:mergereff}, which demonstrates that a ten percent fraction of events coming from GCs could explain the shape of the BH mass spectrum, in addition to the observed merger rate.
Our results suggest that GW clustering measures could provide an independent means to test whether this scenario is the one at play.

In conclusion, we foresee that new types of analysis such as the one presented in this work will become increasingly crucial in the years to come. 
These alternative approaches will not only take advantage of the quality of the data that future observatories will provide, but also help model-builders in constraining the extremely large parameter space that lives behind any model of BBH formation.

%%%%%%%%%%%%%%%%%%%%%%%%%%%%%%%%%%%%%%%%%%%%%%%%%%%%%%%%%%%%%%%%%%%%%%%%%%%%%%%%%%%%%%%%%%%%%%%%%%%%%%%%%%%%%%%%%%%%%%%%%%%%%%%%%%%%

\acknowledgments
The authors thank Lumen Boco for helpful discussions.
NB acknowledges support from the European Union's Horizon Europe research and innovation program under the Marie Sk\l{}odowska-Curie grant agreement no. 101207487 (GWSKY - Mapping the Universe with Gravitational Waves).
MB acknowledges that this article was produced while attending the PhD program in PhD in Space Science and Technology at the University of Trento, Cycle XXXIX, with the support of a scholarship financed by the Ministerial Decree no. 118 of 2nd March 2023, based on the NRRP - funded by the European Union - NextGenerationEU - Mission 4 "Education and Research", Component 1 "Enhancement of the offer of educational services: from nurseries to universities” - Investment 4.1 “Extension of the number of research doctorates and innovative doctorates for public administration and cultural heritage” - CUP E66E23000110001 and support by the Italian grant Project SPACE-IT-UP by the Italian Space Agency and Ministry of University and Research, Contract Number 2024-5-E.0. 
ST acknowledges financial support from the Alexander von Humboldt Foundation for the Humboldt Research Fellowship.
The authors acknowledge support from the European Research Council for the ERC Consolidator grant DEMOBLACK, under contract no. 770017 (PI: M. Mapelli), and for the ERC Advanced grant IMBLACK, under contract no. 101197608 (PI: M. Mapelli).

%%%%%%%%%%%%%%%%%%%%%%%%%%%%%%%%%%%%%%%%%%%%%%%%%%%%%%%%%%%%%%%%%%%%%%%%%%%%%%%%%%%%%%%%%%%%%%%%%%%%%%%%%%%%%%%%%%%%%%%%%%%%%%%%%%%%

\appendix
\include{Appendix}

%%%%%%%%%%%%%%%%%%%%%%%%%%%%%%%%%%%%%%%%%%%%%%%%%%%%%%%%%%%%%%%%%%%%%%%%%%%%%%%%%%%%%%%%%%%%%%%%%%%%%%%%%%%%%%%%%%%%%%%%%%%%%%%%%%%%

\bibliography{bibliography}
\bibliographystyle{utcaps}

\end{document}

%% file: Appendix.tex
\section{Properties of the large-scale structure tracers}
\label{app:tracer_population}

\begin{table}[ht]
    \centerline{
    \begin{tabular}{|c|cccc|}
        \hline
        Dataset & \# bins & $\ell_{\min}$ & $\ell_{\max}$ & Redshift mean~$z_j$ \\
        \hline
        \hline
        Galaxy sample & 10 & 2 & 50/100/200  & 0.4, 0.8, 1.2, 1.6, 2.0, 2.4, 2.8, 3.2, 3.6, 4.0 \\
        GW sample & 7 & 2 & 50/100/200 & 0.4, 0.8, 1.2, 1.6, 2.0, 2.4, 2.9 \\
        \hline
    \end{tabular}}
    \caption{Redshift binning and multipole specifications for both the galaxy and GW samples.
    The values of~$\ell_\mathrm{max}=50,100,200$ refer to the scenarios where GW average angular uncertainty is~$\Delta\Omega = 1, 4, 16\ \mathrm{deg^2}$, respectively, as described in appendix~\ref{subapp:gw_sample}.}
    \label{tab:tracers_binning_properties}
\end{table}

%%%%%%%%%%%%%%%%%%%%%%%%%%%%%%%%%%%%%%%%%%%%%%%%%%%%%%%%%%%%%%%%%%%%%%%%%%%%%%%%%%%%%%%%%%%%%%%%%%%%%%%%%%%%%%%%%%%%%%%%%%%%%%%%%%%%

\subsection{Galaxy sample}

As anticipated in section~\ref{sec:gw_clustering}, in this work we consider a strawman galaxy sample with characteristic number density and bias functions typical of a Stage IV galaxy survey, as in Refs.~\cite{bosi:gwxlssI, scarpel:gwxlss}.
This galaxy population is used either as a star-forming or a quiescent galaxy population to test the impact of galaxy-GW cross-shot noise, while keeping fixed all other specifications.
The galaxy redshift distribution is parametrized as
\begin{equation}
    \frac{d^2N_\mathrm{gal}}{dzd\Omega} = \mathcal{A}_\mathrm{gal} \left( \frac{z}{z_\mathrm{gal}} \right)^{\alpha_\mathrm{gal}} e^{-(z/z_\mathrm{gal})^{\beta_\mathrm{gal}}},
\label{eq:galaxy_dNdzdOmega}
\end{equation}
where~$\left\lbrace \mathcal{A}_\mathrm{gal}, z_\mathrm{gal}, \alpha_\mathrm{gal}, \beta_\mathrm{gal} \right\rbrace = \left\lbrace 25509\ \mathrm{gal/deg^2}, 0.09, 1.75, 0.69 \right\rbrace$.
We divide the galaxies into ten redshift bins reported in table~\ref{tab:tracers_binning_properties}, each one characterized by half-width of~$\Delta z=0.2$.
In each redshift bin, we consider a top-hat window function.
The galaxy bias reads as
\begin{equation}
    b_\mathrm{gal} = b_\mathrm{0,gal} + b_\mathrm{1,gal}z + b_\mathrm{2,gal} z^2,
\end{equation}
where~$\{ b_{0,\mathrm{gal}}, b_{1,\mathrm{gal}}, b_{2,\mathrm{gal}} \}=\{0.53, 1.59, -0.08\}$, the magnification bias is set to~$s_\mathrm{gal} = 0.6$, while the evolution bias is computed directly from the galaxy redshift distribution.

For the purpose of this analysis, it is not possible to take full-advantage of the large angular resolution of galaxy surveys.
Therefore, only multipoles where we can have simultaneously access to galaxy and GW angular power spectra are used in the analysis of sections~\ref{subsec:model_agnostic_estimate} and~\ref{subsec:multi_population_scenario}, as we also report in table~\ref{tab:tracers_binning_properties}.
Additionally, since GW observatories are sensitive to the entire sky, we consider a full-sky survey, i.e.,~$f_\mathrm{sky}=0.7$, to maximize the range of observable multipoles.
Finally, the only source of noise for galaxies is the shot-noise given by 
\begin{equation}
    N^\mathrm{gal,gal}_\ell (z_i,z_j) = \delta^K_{ij} \left( \frac{dN_\mathrm{gal}(z_i)}{d\Omega} \right)^{-1},
\end{equation}
where~$\delta^K_{ij}$ is the Kronecker delta.

%%%%%%%%%%%%%%%%%%%%%%%%%%%%%%%%%%%%%%%%%%%%%%%%%%%%%%%%%%%%%%%%%%%%%%%%%%%%%%%%%%%%%%%%%%%%%%%%%%%%%%%%%%%%%%%%%%%%%%%%%%%%%%%%%%%%

\subsection{Gravitational wave sample}
\label{subapp:gw_sample}

Sections~\ref{subsec:model_agnostic_estimate} and~\ref{subsec:multi_population_scenario} assume different astrophysical models to describe the GW sample, each one of them with its own number density and bias.
However, certain aspects of the analysis are shared between all cases.
One of them is the binning strategy reported in table~\ref{tab:tracers_binning_properties}, i.e., the total number of redshift bins, the redshift means and half-width, and the choice of using Gaussian window functions because of redshift uncertainties, see also Refs.~\cite{bosi:gwxlssI, scarpel:gwxlss}.
Although we are considering events detected by a third generation GW observatory network, we limit our GW sample to redshift~$z\lesssim 3.0$, since for larger redshift the number of events rapidly decreases.
A second common aspect is the total number of GW events used in each~$\mathrm{SNR}^2$ analysis, which is approximately~$N^\mathrm{tot}_\mathrm{GW} \approx 380\ 000$, which is easily achievable for a third generation detector network.

The angular uncertainty of GW events remains a major limitation for cross-correlation analysis.
In particular, the sky localization of individual events ranges between~$\Delta\Omega \in [10^{-3}, 10^2]\ \mathrm{deg^2}$~\cite{iacovelli:gwlocalization} for the ET2CE network, with an average of a few square degrees.
We account for this limitation by restricting the analysis to multipoles~$\ell \leq \ell_\mathrm{max} \simeq 180^\circ / \sqrt{\Delta\Omega}$.
In this work, we choose the three representative values of~$\Delta\Omega = 1, 4, 16\ \mathrm{deg^2}$ that correspond to~$\ell_\mathrm{max} = 200, 100, 50$, respectively.

As for any discrete tracer, GWs are also affected by a shot noise of the form
\begin{equation}
    N_\ell^\mathrm{GW,GW}(z_i, z_j) = \left( \frac{dN_\mathrm{GW}^\mathrm{obs}}{d\Omega} \right)^{-1} \exp\left[ \frac{\ell (\ell+1)\Delta\Omega}{8\log2} \right] \delta^K_{ij},
\end{equation}
where the observed number of GW,~$dN_\mathrm{GW}^\mathrm{obs}/d\Omega$, also accounts for redshift uncertainties, as explained in Refs.~\cite{bosi:gwxlssI, scarpel:gwxlss}, and we account for angular uncertainties with a beam-smearing factor.
Finally, as discussed in section~\ref{sec:gw_clustering}, the galaxy-GW angular power spectrum receives an additional cross-shot noise contribution of the form
\begin{equation}
    N_\ell^\mathrm{gal,GW}(z_i, z_j) = f_\mathrm{GW} \left( \frac{dN_\mathrm{gal}}{d\Omega} \right)^{-1} \exp\left[ \frac{\ell (\ell+1)\Delta\Omega}{16\log2} \right] \delta^K_{ij},
\end{equation}
where~$f_\mathrm{GW}$ is the fraction of observed GW events that is statistically correlated with the galaxy population targeted by the survey.

%%%%%%%%%%%%%%%%%%%%%%%%%%%%%%%%%%%%%%%%%%%%%%%%%%%%%%%%%%%%%%%%%%%%%%%%%%%%%%%%%%%%%%%%%%%%%%%%%%%%%%%%%%%%%%%%%%%%%%%%%%%%%%%%%%%%

\section{Constructing a synthetic Universe}
\label{app:synthetic_universe}

In this appendix, we detail how LSS can be efficiently modeled via a statistical ``Halo Occupation Distribution'' (HOD) approach calibrated both on numerical N-body simulations and observations~\cite{zheng:hod}.
In particular, we are interested in describing how galaxies, i.e., the hosts of GW events, populate dark matter halos. 
The HOD approach can be further extended to also describe how merging BBHs populate galaxies, see, e.g., Refs.~\cite{libanore:gwxlssI, bellomo:classgwb}.

%%%%%%%%%%%%%%%%%%%%%%%%%%%%%%%%%%%%%%%%%%%%%%%%%%%%%%%%%%%%%%%%%%%%%%%%%%%%%%%%%%%%%%%%%%%%%%%%%%%%%%%%%%%%%%%%%%%%%%%%%%%%%%%%%%%%

\subsection{Modeling the Large-Scale Structure of the Universe}

The first step regards developing an analytical description of LSS by extending the approach of Refs.~\cite{rodriguezpuebla:galaxyhodI, rodriguezpuebla:galaxyhodII}.
Galaxies are conveniently divided into two categories, central and satellites.
The number density of central galaxies is
\begin{equation}
    \frac{dn_\mathrm{cen}}{dM_\star} = \int dM_h \frac{dn^\mathrm{cen}_h}{dM_h} p_\mathrm{cen}(M_\star|M_h),
\label{eq:ngal_central}
\end{equation}
where~$dn^\mathrm{cen}_h/dM_h$ is the number density of halos hosting a central galaxy and~$p_\mathrm{cen}$ is the pdf of having a central galaxy of mass~$M_\star$ in a halo of mass~$M_h$.
N-body simulations are typically populated assuming that for each halo the number of central galaxies is extracted from a binomial distribution,
\begin{equation}
    N_\mathrm{cen} \sim \mathrm{B}(1,\mu_\mathrm{cen}),
\end{equation}
where the probability of having a central galaxy is
\begin{equation}
	\mu_\mathrm{cen} = \frac{1}{2} \left[ 1 + \mathrm{erf}\left( 
	\frac{\log (M_h/M^\mathrm{min}_h)}{\sigma_{M_h}} \right) \right],
\end{equation}
where~$M_h^{\rm min}$ and~$\sigma_{M_h}$ are the average minimum halo mass that hosts a central galaxy and its dispersion, respectively.
The average occupation number, i.e., the expected number of central galaxies per halo, is
\begin{equation}
   \left\langle N_\mathrm{cen/h} | M_h \right\rangle = \sum_{N_\mathrm{cen}=0}^{1} N_\mathrm{cen} \binom{1}{N_\mathrm{cen}} \mu_\mathrm{cen}^{N_\mathrm{cen}} (1-\mu_\mathrm{cen})^{1-N_\mathrm{cen}} = \mu_\mathrm{cen},
\end{equation}
so that, in the end, 
\begin{equation}
	\frac{dn^\mathrm{cen}_h}{dM_h} = \frac{dn_h}{dM_h} \mu_\mathrm{cen}.
\end{equation}

On the other hand, the \textit{average} number of satellite galaxies of mass~$M_\star$ in a halo of mass~$M_h$ reads as
\begin{equation}
        \frac{dN_\mathrm{sat}(M_\star|M_h)}{dM_\star} = \sum_{N_\mathrm{sat}=0}^{N_\mathrm{subh}(M_h)} p(N_\mathrm{sat}) \int dM_\mathrm{subh} \frac{dN^\mathrm{sat}_\mathrm{subh}(M_\mathrm{subh}|M_h)}{dM_\mathrm{subh}} p_\mathrm{sat}(M_\star|M_\mathrm{subh}),
\end{equation}
where~$dN^\mathrm{sat}_\mathrm{subh}/dM_\mathrm{subh}$ is the number of subhalos of mass~$M_\mathrm{subh}$ in a halo of mass~$M_h$ hosting a satellite galaxy, $p(N_\mathrm{sat})$ is the probability of having~$N_\mathrm{sat}$ satellite galaxies, $p_\mathrm{sat}$ is the probability that a subhalo of mass~$M_\mathrm{subh}$ hosts a galaxy of mass~$M_\star$, and the latter is assumed to be independent of the halo mass.
The standard picture is that satellites live in subhalos massive enough to host a galaxy, i.e., with masses~$M_\mathrm{subh} \geq M^\mathrm{min}_\mathrm{subh}$, and their number is indicated by
\begin{equation}
	N_\mathrm{subh}(M_h) = \int_{M^\mathrm{min}_\mathrm{subh}}^{M_h}  dM_\mathrm{subh} \frac{dN_\mathrm{subh}(M_\mathrm{subh}|M_h)}{dM_\mathrm{subh}}.
\label{eq:maximum_subhalo_number}
\end{equation}
The number of satellites is typically drawn from a Poissonian distribution as
\begin{equation}
	N_\mathrm{sat} \sim \mathrm{Pois}(\mu_\mathrm{sat}),
\end{equation}
where, provided that the halo already has a central galaxy, the mean number of satellites is given by the parametric form
\begin{equation}
	\mu_\mathrm{sat} = \left( \frac{M_h - M_\mathrm{cut}}{M_1} \right)^{\alpha_\mathrm{sat}},
\end{equation}
or, said otherwise, the average occupation number of satellite galaxies per halo is
\begin{equation}
    \left\langle N_\mathrm{sat/h} | M_h \right\rangle = \sum_{N_\mathrm{sat}=0}^{\infty} N_\mathrm{sat} p(N_\mathrm{sat}) = \sum_{N_\mathrm{sat}=0}^{\infty} N_\mathrm{sat} \frac{\mu_\mathrm{sat}^{N_\mathrm{sat}} e^{-\mu_\mathrm{sat}}}{N_\mathrm{sat}!} = \mu_\mathrm{sat}.
\end{equation}
However, since in this picture satellites can only live in subhalos, the pdf of the number of satellites cannot be exactly Poissonian because the maximum number of satellites is given by the~$N_\mathrm{subh}$ upper bound in equation~\eqref{eq:maximum_subhalo_number}.
Nevertheless, as long as~$\mu_\mathrm{sat} \ll N_\mathrm{subh}$, this approximation holds without introducing any significant error.

On the other hand, estimating the subhalo occupation number requires some additional consideration.
Let us consider the ``simplified'' case where subhalo masses can only take discrete values~$\{M_1, ..., M_M\}$, which in a given halo appear with different frequencies~$\{ K_1, ... , K_M \}$.
The number of subhalos is given by~$N_\mathrm{subh} = \sum^M_{j=1} K_j$, and our goal is to randomly draw (without replacement)~$N_\mathrm{sat} \leq N_\mathrm{subh}$ subhalos that will host a satellite galaxy. 
The probability of having~$k_1\leq K_1$ subhalos with mass~$M_1$ hosting a satellite is given by the hypergeometric distribution, which reads as
\begin{equation}
	p(k_1) = \frac{\binom{K_1}{k_1} \binom{N_\mathrm{subh}-K_1}{N_\mathrm{sat}-k_1}}{\binom{N_\mathrm{subh}}{N_\mathrm{sat}}}.
\end{equation}
Therefore, the average satellite occupation number of subhalos with mass~$M_1$ is
\begin{equation}
	N_1 = \sum_{k_1=0}^{K_1} k_1 p(k_1) = N_\mathrm{sat} \frac{K_1}{N_\mathrm{subh}}.
\end{equation}
If we now generalize this result to the continuum limit of masses, we have
\begin{equation}
	\frac{dN^\mathrm{sat}_\mathrm{subh}}{dM_\mathrm{subh}} = \frac{N_\mathrm{sat}}{N_\mathrm{subh}} \frac{dN_\mathrm{subh}}{dM_\mathrm{subh} },
\end{equation}
and the average number of satellite galaxies becomes
\begin{equation}
    \frac{dN_\mathrm{sat}(M_\star|M_h)}{dM_\star} = \frac{\mu_\mathrm{sat}}{N_\mathrm{subh}} \int dM_\mathrm{subh} \frac{dN_\mathrm{subh}}{dM_\mathrm{subh} } p_\mathrm{sat}(M_\star|M_\mathrm{subh}).
\label{eq:average_satellite_number}
\end{equation}

At this point, similarly to equation~\eqref{eq:ngal_central}, the number density of satellite galaxies is given by
\begin{equation}
    \frac{dn_\mathrm{sat}}{dM_\star} = \int dM_h \frac{dn_h}{dM_h} \frac{dN_\mathrm{sat}(M_\star|M_h)}{dM_\star} = \int dM_h dM_\mathrm{subh} \frac{\mu_\mathrm{sat}}{N_\mathrm{subh}} \frac{d^2n_\mathrm{subh}}{dM_\mathrm{subh}dM_h} p_\mathrm{sat}(M_\star|M_\mathrm{subh}),
\end{equation}
where, by definition, 
\begin{equation}
    \frac{d^2n_\mathrm{subh}}{dM_\mathrm{subh} dM_h} = \frac{dn_h}{dM_h} \frac{dN_\mathrm{subh}}{dM_\mathrm{subh}}.
\end{equation}
In general, this formalism predicts that the average galaxy occupation number per halo is
\begin{equation}
    \begin{aligned}
        \left\langle \frac{dN_\mathrm{gal/h}}{dM_\star} \bigg| M_h \right\rangle &= \left\langle \frac{dN_\mathrm{cen/h}}{dM_\star} \bigg| M_h \right\rangle + \left\langle \frac{dN_\mathrm{sat/h}}{dM_\star} \bigg| M_h \right\rangle \\
        &\qquad = \frac{d^2n_\mathrm{cen}}{dM_\star dM_h} \bigg/ \frac{dn_h}{dM_h} + \frac{d^2n_\mathrm{sat}}{dM_\star dM_h} \bigg/ \frac{dn_h}{dM_h} \\
        &\qquad = \mu_\mathrm{cen} p_\mathrm{cen}(M_\star|M_h) + \frac{\mu_\mathrm{sat}}{N_\mathrm{subh}} \int dM_\mathrm{subh} \frac{dN_\mathrm{subh}}{dM_\mathrm{subh}} p_\mathrm{sat}(M_\star|M_\mathrm{subh}),
    \end{aligned}
\end{equation}
where we note that by integrating over the galaxy stellar mass we self-consistently recover the expected total number of galaxies per halo.

%%%%%%%%%%%%%%%%%%%%%%%%%%%%%%%%%%%%%%%%%%%%%%%%%%%%%%%%%%%%%%%%%%%%%%%%%%%%%%%%%%%%%%%%%%%%%%%%%%%%%%%%%%%%%%%%%%%%%%%%%%%%%%%%%%%%

\subsection{Abundance matching scheme}

The formalism outlined above is still compatible with the simplest versions of the widely-used abundance matching technique, which is based on the assumption that to each halo/subhalo corresponds a central/satellite galaxy, i.e.,~$\mu_\mathrm{cen}=1$ and~$p(N_\mathrm{sat})=\delta^K_{N_\mathrm{sat},N_\mathrm{subh}}$, in such a way that
\begin{equation}
	\frac{dN^\mathrm{sat}_\mathrm{subh}}{dM_\mathrm{subh}} = \frac{dN_\mathrm{subh}}{dM_\mathrm{subh}}.
\end{equation}
In this context, the probability describing galaxy masses are typically assumed to be
\begin{equation}
        p_\mathrm{cen}(M_\star|M_h) = \delta^D\left[M_\star - M^\mathrm{cen}_\star(M_h)\right], \qquad p_\mathrm{sat}(M_\star|M_\mathrm{subh}) = \delta^D\left[M_\star - M^\mathrm{sat}_\star(M_\mathrm{subh}) \right],
\end{equation}
where~$\{ M^\mathrm{cen}_\star(M_h), M^\mathrm{sat}_\star(M_\mathrm{subh}) \}$ are the stellar-mass-to-halo/subhalo-mass functions for central/satellite galaxies, respectively, and the cumulative number density of galaxies of mass above a given value~$\overline{M}_\star$ is chosen to be in one-to-one correspondence with the halo/subhalo number densities.
At the practical level, we have
\begin{equation}
    \begin{aligned}
        & n_\mathrm{gal}(M_\star\geq \overline{M}_\star) = n_\mathrm{cen}(M_\star\geq \overline{M}_\star) + n_\mathrm{sat}(M_\star\geq \overline{M}_\star) \\
        &\quad = \int^\infty_{\overline{M}_\star} dM_\star dM_h \frac{dn_h}{dM_h}  \delta^D\left[M_\star - M^\mathrm{cen}_\star(M_h)\right] + \int^\infty_{\overline{M}_\star} dM_\star  dM_\mathrm{subh} \frac{dn_\mathrm{subh}}{dM_\mathrm{subh}} \delta^D\left[M_\star - M^\mathrm{sat}_\star(M_\mathrm{subh}) \right] \\
        &\quad = \int dM_h \frac{dn_h}{dM_h} \Theta_H\left[M^\mathrm{cen}_\star(M_h) - \overline{M}_\star \right] + \int dM_\mathrm{subh} \frac{dn_\mathrm{subh}}{dM_\mathrm{subh}} \Theta_H\left[ M^\mathrm{sat}_\star(M_\mathrm{subh}) - \overline{M}_\star \right] \\
        &\quad = \int dM_h \frac{dn_h}{dM_h} \Theta_H \left[M_h - M^\mathrm{cen}_h (\overline{M}_\star) \right] + \int dM_\mathrm{subh} \frac{dn_\mathrm{subh}}{dM_\mathrm{subh}} \Theta_H \left[ M_\mathrm{subh} - M^\mathrm{sat}_\mathrm{subh} (\overline{M}_\star) \right] \\
        &\quad = n_h \left(M_h \geq M^\mathrm{cen}_h (\overline{M}_\star) \right) + n_\mathrm{subh} \left( M_\mathrm{subh} \geq M^\mathrm{sat}_\mathrm{subh} (\overline{M}_\star) \right),
    \end{aligned}
\end{equation}
where, in the end, the matching is done separately for the halo/central and subhalo/satellite pairs.

The procedure described above is certainly useful when an N-body simulation is available; however, that is not the case in our analytical approach. 
In this work, we take a slightly more general approach, which allows for \textit{(i)} the possibility of having subhalos not necessarily hosting a satellite galaxy, i.e.,~$N_\mathrm{sat} < N_\mathrm{subh}$, and \textit{(ii)} the introduction of some additional intrinsic scattering into the characterization of GW hosts focusing, in particular, on galaxy masses.
The central and satellite galaxy pdfs determine the properties of the galaxy population, since halo and subhalo mass functions can be derived from matter-only N-body simulation for each cosmology.
A common choice is to consider lognormal distributions for both pdfs.
As commented before, each population is characterized by its own mean~$\{ M^\mathrm{cen}_\star(M_h), M^\mathrm{sat}_\star(M_\mathrm{subh}) \}$ and, additionally, by a dispersion~$\{\sigma_\mathrm{cen}, \sigma_\mathrm{sat} \}$, in such a way that
\begin{equation}
	p_\mathrm{cen}(M_\star|M_\mathrm{h}) = \frac{e^{-\frac{\log^2(M_\star/M^\mathrm{cen}_\star(M_h))}{2\sigma^2_\mathrm{cen}}}}{\sqrt{2\pi}\sigma_\mathrm{cen} M_\star}, \qquad p_\mathrm{sat}(M_\star|M_\mathrm{subh}) = \frac{e^{-\frac{\log^2(M_\star/M^\mathrm{sat}_\star(M_\mathrm{subh}))}{2\sigma^2_\mathrm{sat}}}}{\sqrt{2\pi}\sigma_\mathrm{sat} M_\star}.
\end{equation}

%%%%%%%%%%%%%%%%%%%%%%%%%%%%%%%%%%%%%%%%%%%%%%%%%%%%%%%%%%%%%%%%%%%%%%%%%%%%%%%%%%%%%%%%%%%%%%%%%%%%%%%%%%%%%%%%%%%%%%%%%%%%%%%%%%%%

\subsection{Galaxy bias and its implementation}

At this point, we can actively compute any galaxy population bias using the HOD approach.
In particular, the number density and galaxy bias are given by
\begin{equation}
	\frac{dn_\mathrm{gal}}{dM_\star} = \int dM_h \frac{dn_h}{dM_h} \left\langle \frac{dN_\mathrm{gal/h}}{dM_\star} \bigg| M_h \right\rangle, \qquad \frac{dn_\mathrm{gal}}{dM_\star} b_\mathrm{gal} = \int dM_h \frac{dn_h}{dM_h} \left\langle \frac{dN_\mathrm{gal/h}}{dM_\star} \bigg| M_h \right\rangle b_h,
\end{equation}
where~$b_h$ is the halo bias.
Sometimes it is more convenient to consider a galaxy mass bin centered around some mean value~$\overline{M}_\star$.
In that case, the galaxy number density and bias are recast in terms of
\begin{equation}
    \begin{aligned}
        n_\mathrm{gal}(\overline{M}_\star) &= \int dM_h \frac{dn_h}{dM_h} \left\langle N_\mathrm{gal/h}(\overline{M}_\star) | M_h \right\rangle, \\
        b_\mathrm{gal}(\overline{M}_\star) &= n^{-1}_\mathrm{gal}(\overline{M}_\star) \int dM_h \frac{dn_h}{dM_h} \left\langle N_\mathrm{gal/h}(\overline{M}_\star) | M_h \right\rangle b_h,	
    \end{aligned}
\end{equation}
where the average occupation number in the mass bin is
\begin{equation}
	\left\langle N_\mathrm{gal/h}(\overline{M}_\star) | M_h \right\rangle = \int_{\overline{M}_\star} dM_\star \left\langle \frac{dN_\mathrm{gal/h}}{dM_\star} \bigg| M_h \right\rangle.
\end{equation}

In this work, we use the halo mass function of Ref.~\cite{rodriguezpuebla:halomassfunction}, and consider a population of dark matter halos in the mass range of~$M_h \in [10^{10}, M_h^\mathrm{max}(z)]$.
The redshift-dependent maximum halo mass is given by~$\log_{10} M_h^\mathrm{max}(z) = 13.54-0.24z+2.02e^{-z/4.48}$, and corresponds to~$M_h^\mathrm{max} = 3.6\times 10^{15}\ M_\odot$ at redshift~$z=0$~\cite{behroozi:universemachine}.
The chosen values of the average minimum halo mass to host a central galaxy and its dispersion are~$M_h^\mathrm{min}=10^{10}\ M_\odot$ and~$\sigma_{M_h}=0.7$.
The subhalo mass function is also taken from Ref.~\cite{rodriguezpuebla:halomassfunction}, and reads as
\begin{equation}
    \frac{dN_\mathrm{subh}}{dM_\mathrm{subh}} = -\frac{d\mu}{dM_\mathrm{subh}} \frac{d}{d\mu} \left[ \mu_0 \left( \frac{\mu}{\mu_1} \right)^\alpha e^{-(\mu/\mu_\mathrm{cut})^\beta} \right],
\end{equation}
where~$\mu=M_\mathrm{subh}/M_h$ and~$\mu_0 = \left[ M_h/(10^{12}\ M_\odot) \right]^\gamma$.
Since dark matter is easily stripped from subhalos after mergers, the mass of satellite galaxies is more correlated to the subhalo mass at the time of accretion into the parent halo than at the time of observation.
Therefore, in terms of the subhalo mass, we consider~$M_\mathrm{subh}\equiv M_\mathrm{acc}\in [M^\mathrm{min}_\mathrm{subh}, M_h/2]\ M_\odot$, where the minimum subhalo mass is~$M^\mathrm{min}_\mathrm{subh} = 10^9\ M_\odot$, and fix~$\mu_1 = 0.030$, $\mu_\mathrm{cut} = 0.199$, $\alpha = -0.777$, $\beta = 1.210$, and~$\gamma = 0.102$. 
The chosen values of the parameters appearing in the mean number of satellites are~$M_\mathrm{cut} = 0\ M_\odot$, $M_1 = 10^{11}\ M_\odot$, and~$\alpha_\mathrm{sat} = 1$ in the redshift range of interest for this work.
Finally, the halo bias is taken from Ref.~\cite{tinker:halobias}.

Regarding the central/satellite pdfs, the mean value of galaxy masses are selected using the stellar-mass-to-halo-mass relation from Ref.~\cite{behroozi:universemachine}.
Both pdfs are expected to be relatively narrow, with dispersions of the order of~$\sigma_\mathrm{cen},\sigma_\mathrm{sat} \approx 0.5-0.7$~\cite{rodriguezpuebla:galaxyhodII, behroozi:universemachine}.
In the case of subhalos, we evaluate the stellar-mass-to-halo-mass relation at the subhalo accretion redshift~$z_\mathrm{acc}$.
However, in order to account for the mismatch between the redshift of observation and accretion, we have to slightly modify equation~\eqref{eq:average_satellite_number} as
\begin{equation}
    \frac{dN_\mathrm{sat}(M_\star|M_h,z)}{dM_\star} = \frac{\mu_\mathrm{sat}}{N_\mathrm{subh}} \int dM_\mathrm{subh} \frac{dN_\mathrm{subh}}{dM_\mathrm{subh}} \int_{\mathrm{max}(z,z^\mathrm{min}_\mathrm{acc})}^{z^\mathrm{max}_\mathrm{acc}} dz_\mathrm{acc} p(z_\mathrm{acc}|M_h,M_\mathrm{sub}) p_\mathrm{sat}(M_\star|M_\mathrm{subh}, z_\mathrm{acc}),
\end{equation}
where~$p(z_\mathrm{acc}|M_h,M_\mathrm{sub})$ is the probability of accreting a subhalo of mass~$M_\mathrm{subh}$ into a halo of mass~$M_h$ at redshift~$z_\mathrm{acc}$.
The rest of the derivation proceeds as explained above.

The accretion redshift pdf is taken to be proportional to the derivative with respect to the redshift of the halo
virial mass growth, i.e., $p(z_\mathrm{acc}) \propto dM_h/dz$, see, e.g., Ref.~\cite{behroozi:halomassgrowth}.
The minimum and maximum accretion masses bracket the redshift interval in which the halo accretes at least as much mass as the subhalo has in a time smaller than the halo dynamical time~$t_\mathrm{dyn}(z)$, to differentiate between a merger events and the natural growth due to matter infall.
In practice, the halo dynamical time approximately corresponds to a redshift interval of~$\Delta z_\mathrm{dyn}(z) \approx (1+z)t_\mathrm{dyn}(z)/H(z) = (1+z) \sqrt{8\pi / 3\Delta_\mathrm{vir}}$~\cite{rodriguezpuebla:halomassfunction}, where~$\Delta_\mathrm{vir}$ is the mean halo overdensity~\cite{bryan:halooverdensity}.
We fix a redshift interval~$\Delta z_\mathrm{acc} = \Delta z_\mathrm{dyn}/10 \ll \Delta z_\mathrm{dyn}$, and compute the minimum and maximum accretion redshift as the bounds of the redshift interval where~$M_h(z_\mathrm{acc})-M_h(z_\mathrm{acc}+\Delta z_\mathrm{acc}) \geq M_\mathrm{subh}$, if such a condition can be satisfied.
If the condition cannot be satisfied, the pdf is identically zero.
In the end, in our synthetic Universe, typical galaxies have masses in the range~$M_\star \in [10^{6.1}, 10^{11.7}]\ M_\odot$.

%%%%%%%%%%%%%%%%%%%%%%%%%%%%%%%%%%%%%%%%%%%%%%%%%%%%%%%%%%%%%%%%%%%%%%%%%%%%%%%%%%%%%%%%%%%%%%%%%%%%%%%%%%%%%%%%%%%%%%%%%%%%%%%%%%%%

\subsection{Subgalactic structure}

Given the goal of our analysis, we need to further characterize the subgalactic environment, in particular Globular Clusters (GC).
GCs are a common kind of structure not only in our Milky Way, but also across galaxies in the entire Universe. 
Here, we are not interested in describing their internal structure, but only their global properties.
Following Ref.~\cite{mapelli:fastclusterII}, we assume that the GC population is characterized by a lognormal mass distribution with mean~$\mu_\mathrm{GC} = 10^{5.9}\ M_\odot$ and dispersion~$\sigma_\mathrm{GC} = 0.9$.
Therefore, in this mass range, the mass function reads as
\begin{equation}
	\frac{dN_\mathrm{GC}(M_\mathrm{GC}|M_h)}{dM_\mathrm{GC}} = \frac{2 N^\mathrm{tot}_\mathrm{GC/h} }{ \mathrm{erf} \left[ \frac{\log(M^\mathrm{max}_\mathrm{GC} / \mu_\mathrm{GC})}{\sqrt{2}\sigma_\mathrm{GC}} \right] - \mathrm{erf} \left[ \frac{\log(M^\mathrm{min}_\mathrm{GC} / \mu_\mathrm{GC})}{\sqrt{2}\sigma_\mathrm{GC}} \right]} \frac{e^{-\frac{\log^2 \left( M_\mathrm{GC}/\mu_\mathrm{GC} \right)}{2\sigma^2_\mathrm{GC}}}}{\sqrt{2\pi} \sigma_\mathrm{GC} M_\mathrm{GC}},
\label{eq:GC_occupation_number}
\end{equation}
where the total number of GC per halo is given by
\begin{equation}
	N^\mathrm{tot}_\mathrm{GC/h} = \frac{\eta M_h}{\mu_\mathrm{GC}} e^{-\sigma^2_\mathrm{GC}/2} \frac{\mathrm{erf} \left[ \frac{\log(M^\mathrm{max}_\mathrm{GC} / \mu_\mathrm{GC})}{\sqrt{2}\sigma_\mathrm{GC}} \right] - \mathrm{erf} \left[ \frac{\log(M^\mathrm{min}_\mathrm{GC} / \mu_\mathrm{GC})}{\sqrt{2}\sigma_\mathrm{GC}} \right]}{\mathrm{erf} \left[ \frac{\log(M^\mathrm{max}_\mathrm{GC} e^{-\sigma^2_\mathrm{GC}} / \mu_\mathrm{GC})}{\sqrt{2}\sigma_\mathrm{GC}} \right] - \mathrm{erf} \left[ \frac{\log(M^\mathrm{min}_\mathrm{GC} e^{-\sigma^2_\mathrm{GC}} / \mu_\mathrm{GC})}{\sqrt{2}\sigma_\mathrm{GC}} \right]}.
\end{equation}
The total mass in GCs per halo has been observed to be constant across multiple types of environment, see, e.g., Ref.~\cite{harris:gcmassperhalo} and Refs. therein.
In this work, we assume a fiducial value of~$\eta = M_\mathrm{GC/h}/M_h = 5\times 10^{-5}$.
This additional condition limits the mass range of GC in not very massive halos: the minimum GC mass we consider is approximately~$M^\mathrm{min}_\mathrm{GC} \approx 10^4\ M_\odot$, while the maximum GC is at most~$M^\mathrm{max}_\mathrm{GC} \approx \eta M_h$.
Finally, equation~\eqref{eq:GC_occupation_number} also describes the average occupation number of GC per halo, which does not depend on the stellar masses of the galaxies hosted in the halo.

Thanks to the HOD formalism introduced above, we can easily estimate the GC number density and bias as
\begin{equation}
	\frac{dn_\mathrm{GC}}{dM_\mathrm{GC}} = \int dM_h \frac{dn_h}{dM_h} \frac{dN_\mathrm{GC}}{dM_\mathrm{GC}}, \qquad \frac{dn_\mathrm{GC}}{dM_\mathrm{GC}} b_\mathrm{GC} = \int dM_h \frac{dn_h}{dM_h} \frac{dN_\mathrm{GC}}{dM_\mathrm{GC}} b_h.
\end{equation}
As before, we can also define the same quantities in a mass bin centered at mass~$\overline{M}_\mathrm{GC}$, leading to
\begin{equation}
    \begin{aligned}
        n_\mathrm{GC}(\overline{M}_\mathrm{GC}) &= \int dM_h \frac{dn_h}{dM_h} N_\mathrm{GC}(\overline{M}_\mathrm{GC}), \\
        b_\mathrm{GC}(\overline{M}_\mathrm{GC}) &= n^{-1}_\mathrm{GC}(\overline{M}_\mathrm{GC}) \int dM_h \frac{dn_h}{dM_h} N_\mathrm{GC}(\overline{M}_\mathrm{GC}) b_h,	
    \end{aligned}
\end{equation}
where
\begin{equation}
	N_\mathrm{GC}(\overline{M}_\mathrm{GC}) = \int_{\overline{M}_\mathrm{GC}} dM_\mathrm{GC} \frac{dN_\mathrm{GC}}{dM_\mathrm{GC}}.
\end{equation}

%%%%%%%%%%%%%%%%%%%%%%%%%%%%%%%%%%%%%%%%%%%%%%%%%%%%%%%%%%%%%%%%%%%%%%%%%%%%%%%%%%%%%%%%%%%%%%%%%%%%%%%%%%%%%%%%%%%%%%%%%%%%%%%%%%%%

\subsection{Cosmic star formation}

Once the LSS scheme has been defined, we can start characterizing the environment in which BBHs form and merge.
In particular, we are interested in how many of these binaries form and their dependence on the properties of their host galaxies.
However, there are multiple avenues to infer the cosmic star formation rate density, all calibrated on different observational datasets, with different degrees of refinement.
For the purpose of the agnostic analysis presented in section~\ref{subsec:model_agnostic_estimate}, we adopt the cosmic star formation rate density provided by the \textsc{UniverseMachine} approach~\cite{behroozi:universemachine}, as done for instance in Ref.~\cite{bellomo:classgwb}.
In this case, the cosmic star formation rate density reads as 
\begin{equation}
    \dot{\rho}_\mathrm{CSFR}(z) = \int dM_h d\mathrm{SFR} p(\mathrm{SFR}|M_h,z) \mathrm{SFR} \frac{dn_h}{dM_h},
\end{equation}
where~$\mathrm{SFR}$ is the star-formation rate and~$p(\mathrm{SFR}|M_h,z)$ is the parametric form of the SFR pdf given a dark matter halo of mass~$M_h$ at redshift~$z$ calibrated on a variety of different observations, such as the stellar mass function, quenched fractions, cosmic and specific star formation rates, high-redshift UV luminosity functions, high-redshift UV-stellar mass relations, correlations functions, the dependence of the quenched fractions of central galaxies as a function of the environment and the average infrared excess as a function of the UV luminosity.

On the other hand, for the analysis of section~\ref{subsec:multi_population_scenario} we need a more detailed characterization of the properties of GW hosts.
In that analysis, we start by defining the model for the cosmic star-formation rate density as
\begin{equation}
    \dot{\rho}_\mathrm{CSFR}(z) = \int dZ d\mathrm{SFR} dM_\star \frac{d^3 n_\mathrm{gal}}{dZ d\mathrm{SFR} dM_\star} \mathrm{SFR},
\label{eq:cosmicSFR_def}
\end{equation}
where~$Z$ is the metallicity.
For the purpose at hand, it is more convenient to work in terms of pdfs, as
\begin{equation}
    \frac{d^3 n_\mathrm{gal}}{dZ d\mathrm{SFR} dM_\star} = n_\mathrm{gal}(z) p(Z, \mathrm{SFR}, M_\star| z) = n_\mathrm{gal}(z) p(Z| \mathrm{SFR}, M_\star, z)  p(\mathrm{SFR}| M_\star, z) p(M_\star| z),
\end{equation}
where, for instance,~$p(Z, \mathrm{SFR}, M_\star| z)$ is the conditional pdf of having a galaxy with a given metallicity, star formation rate, and stellar mass at a given redshift.
In this sense, the usual galaxy stellar mass function is simply
\begin{equation}
    \frac{dn_\mathrm{gal}}{dM_\star} = n_\mathrm{gal}(z) p(M_\star|z).
\end{equation}

\begin{figure}
    \centerline{
    \includegraphics[width=\linewidth]{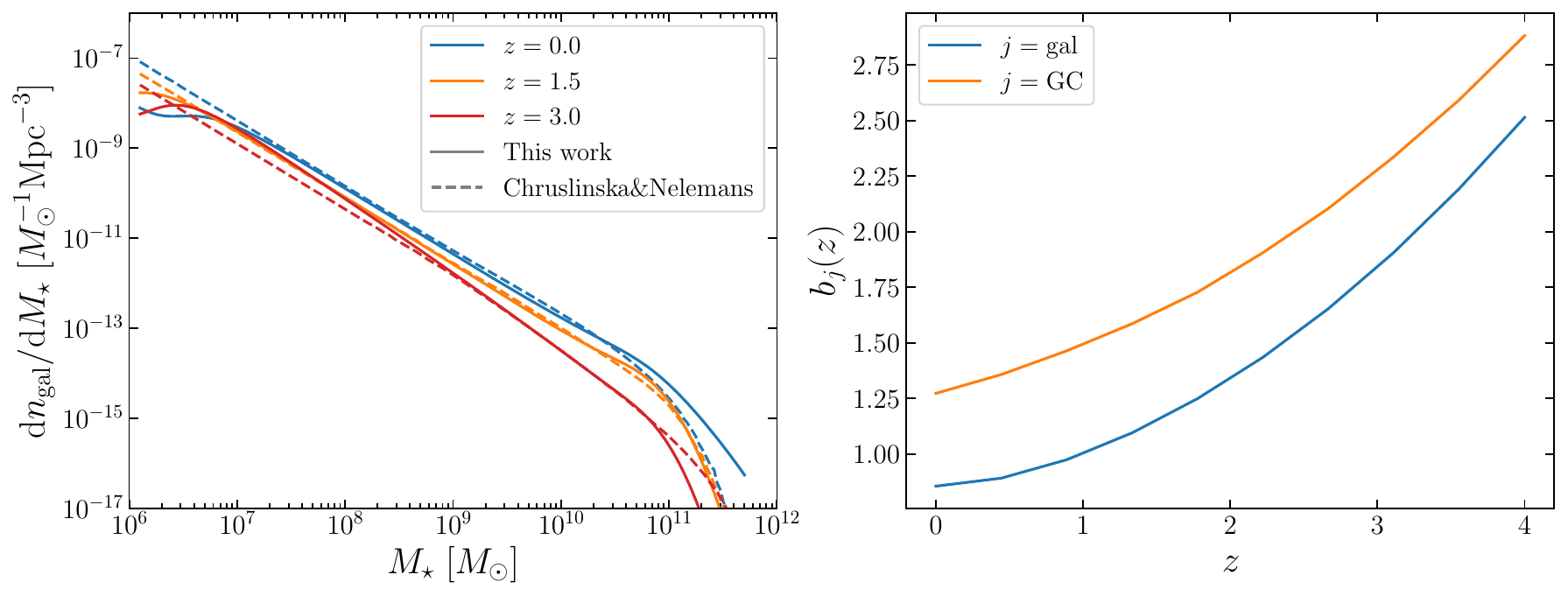}}
    \caption{\textit{Left panel:} comparison between the galaxy stellar mass function obtained from our model with respect to that calibrated on observations provided by Ref.~\cite{chruslinska:gsmf} at different redshifts.
    \textit{Right panel:} galaxy and GC bias obtained with the HOD approach of our model.}
\label{fig:GSMF_bias}
\end{figure}

Although being one of the most commonly investigated properties of galaxy populations, the galaxy stellar mass function lacks  information about dark matter halos that host them.
In this sense, an HOD approach appears quite challenging.
However, thanks to the approach developed in the previous sections, we are able to connect the halo population to the observed galaxy one and to compute both galaxy and GC biases.
As we show in the left panel of figure~\ref{fig:GSMF_bias}, the approach developed in the previous section describes quite accurately widely-used galaxy stellar mass function calibrated on observations such as that of Ref.~\cite{chruslinska:gsmf}.
Note that the agreement is obtained without resorting to any best-fitting strategy; we simply adopted physically motivated values for each of the components entering into the theoretical model.
In the right panel of the same figure, we also show the galaxy and GC bias obtained as explained in the previous sections.
Although different structures are expected to trace LSS in different fashions, this result constitutes the first indication that merging BBHs living in different environments can have different clustering signals.

Regarding metallicities, following Ref.~\cite{boco:metallicitypdf}, we assume a lognormal pdf
\begin{equation}
	p(Z| \mathrm{SFR}, M_\star, z) = \frac{e^{-\log^2\left[ Z / \mu_\mathrm{FMR} \right] / 2\sigma^2_\mathrm{FMR}}}{\sqrt{2\pi} \sigma_\mathrm{FMR} Z},
\end{equation}
where the dispersion is given by~$\sigma_\mathrm{FMR} = 0.23$ and the mean value is given by the fundamental metallicity relation parametrized as in~\cite{chruslinska:fmrI, boco:mr}:
\begin{equation}
    \mu_\mathrm{FMR} = \bar{Z}_\mathrm{FMR}(\mathrm{SFR}, M_\star) = 0.0153 \frac{10^{(Z_{0,\mathrm{FMR}}-8.76)}}{\mathrm{OFe}(\mathrm{sSFR})} \left[ 1 + \left(\frac{M_\star}{M_{0,\mathrm{FMR}}}\right)^{-\beta_\mathrm{FMR}} \right]^{- \frac{\gamma_\mathrm{FMR}}{\beta_\mathrm{FMR}}},
\end{equation}
which also includes the specific star-formation rate ($\mathrm{sSFR}$) oxygen-to-iron correction function~$\mathrm{OFe}$ derived in~\cite{chruslinska:fmrIII, chruslinska:fmrII}.
The fiducial values chosen for our implementation are $Z_{0,\mathrm{FMR}} = 9.0$, $\beta_\mathrm{FMR} = 2.1$,  $\gamma_\mathrm{FMR} = 0.43$, $m_{0,\mathrm{FMR}} = 10.11$,  $\nabla_{0,\mathrm{FMR}} = 0.27$, and
\begin{equation}
	M_{0,\mathrm{FMR}} = 10^{m_{0,\mathrm{FMR}}} \mathrm{SFR}^{\nabla_{0,\mathrm{FMR}}/\gamma_\mathrm{FMR}}.
\end{equation}
The values of~$\beta_\mathrm{FMR}$ and~$m_{0,\mathrm{FMR}}$ are taken from Ref.~\cite{curti:fmr}.
For more details on the fiducial values see~\cite{chruslinska:fmrI, boco:mr}.

The SFG population receives contributions from both main sequence and starburst galaxies; therefore, the global star-formation rate distribution will be modeled as the sum of two contributions. 
Each contribution is described by a lognormal distribution in such a way that the total pdf reads as~\cite{sargent:sfrpdf}
\begin{equation}
    p(\mathrm{SFR}| M_\star, z_f) = f_\mathrm{SB} \frac{e^{-\log^2\left[ \mathrm{SFR} / \mu_\mathrm{SB} \right] / 2\sigma^2_\mathrm{SB}}}{\sqrt{2\pi} \sigma_\mathrm{SB} \mathrm{SFR}} + (1 - f_\mathrm{SB}) \frac{e^{-\log^2\left[ \mathrm{SFR} / \mu_\mathrm{MS} \right] / 2\sigma^2_\mathrm{MS}}}{\sqrt{2\pi} \sigma_\mathrm{MS} \mathrm{SFR}},        
\end{equation}
where the mean values for main sequence and starburst SFR read as~\cite{popesso:sfrms, sgalletta:sfrsb}
\begin{equation}
	\begin{aligned}
		\log_{10}\mu_\mathrm{MS} &= \left(\alpha_{0,\mathrm{MS}} + \alpha_{1,\mathrm{MS}} \log_{10}M_\star \right) t(z) + \beta_{0,\mathrm{MS}} + \beta_{1,\mathrm{MS}} \log_{10}M_\star + \beta_{2,\mathrm{MS}} \log^2_{10}M_\star, \\
		\log_{10}\mu_\mathrm{SB} &= \log_{10}\mu_\mathrm{MS} + 0.59,
	\end{aligned}
\end{equation}
respectively, where~$t(z)$ is the cosmic time, and the dispersions are given by~$\sigma_\mathrm{MS} = 0.432$ and~$\sigma_\mathrm{SB} = 0.560$.
The fiducial values used in our implementation are~$\alpha_{0,\mathrm{MS}} = 2\times 10^{-4}\ \mathrm{Myr}^{-1}$, $\alpha_{1,\mathrm{MS}} = -3.4\times 10^{-5}\ \mathrm{Myr}^{-1}$, $\beta_{0,\mathrm{MS}} = -26.134$, $\beta_{1,\mathrm{MS}} = 4.722$, $\beta_{2,\mathrm{MS}} = -0.1925$.
The values of~$f_\mathrm{SB}$ are taken from appendix~A of Ref.~\cite{chruslinska:fmrII}.

%%%%%%%%%%%%%%%%%%%%%%%%%%%%%%%%%%%%%%%%%%%%%%%%%%%%%%%%%%%%%%%%%%%%%%%%%%%%%%%%%%%%%%%%%%%%%%%%%%%%%%%%%%%%%%%%%%%%%%%%%%%%%%%%%%%%

\subsection{Binary population}
\label{subapp:binary_population}

We assume that, independently of the properties of the galaxy, the mass function of primary companion stars follows a Kroupa distribution~\cite{kroupa:initialstellarmassfunction}
\begin{equation}
	\frac{dN_\star}{dm_\star} = N_\mathrm{bin}\ p(m_1) = \frac{N_\mathrm{bin}}{\mathcal{N}_\star}
	\left\lbrace \begin{aligned}
	& m^{\alpha_1}_1, \quad & m_1 \in [m_\mathrm{min}, m_\mathrm{break}], \\
	& m^{\alpha_2}_1, \quad & m_1 \in [m_\mathrm{break}, m_\mathrm{max}], \\	 
	\end{aligned} \right.
\label{eq:kroupa_imf}
\end{equation}
where~$m_1$ is the mass of the primary star, $N_\mathrm{bin}$ is the total number of binaries, and the normalization factor is given by
\begin{equation}
    \mathcal{N}_\star = \left(\alpha_1 + 1\right)^{-1} \left( m^{\alpha_1 + 1}_\mathrm{break} - m^{\alpha_1 + 1}_\mathrm{min} \right) + \left(\alpha_2 + 1\right)^{-1} \left( m^{\alpha_2 + 1}_\mathrm{max} - m^{\alpha_2 + 1}_\mathrm{break} \right),        
\end{equation}
and the fiducial values of the Kroupa distribution reads are~$\{ m_\mathrm{min}, m_\mathrm{break}, m_\mathrm{max} \} = \{ 0.1, 0.5, 150\}\ M_\odot$ and~$\{\alpha_1, \alpha_2 \} = \{-1.3, -2.3 \}$.
On the other hand, the secondary mass~$m_2$ is sampled from the conditional probability~\cite{sana:secondarymasspdf}
\begin{equation}
    p(m_2|m_1) = \frac{(\alpha_q+1) \left( m_2/m_1 \right)^{\alpha_q}}{m_1 \left[ 1 - \left( m_{2,\mathrm{min}}/m_1 \right)^{\alpha_q+1} \right]},
\end{equation}
where~$m_2 \in [m_{2,\mathrm{min}}, m_1]$, $m_{2,\mathrm{min}} = \mathrm{max}\left[m_\mathrm{min}, m_1/10 \right]$, and~$\alpha_q=-0.1$.
The average mass of the binary system population reads as
\begin{equation}
    \left\langle m_\mathrm{bin} \right\rangle = \int dm_1 dm_2 (m_1 + m_2) p(m_1) p(m_2|m_1) = \left\langle m_1 \right\rangle + \left\langle m_2 \right\rangle,
\end{equation}
where
\begin{equation}
    \begin{aligned}
        \left\langle m_1 \right\rangle &= \mathcal{N}^{-1}_\star \left[ \left(\alpha_1 + 2\right)^{-1} \left( m^{\alpha_1 + 2}_\mathrm{break} - m^{\alpha_1 + 2}_\mathrm{min} \right) \right. + \left. \left(\alpha_2 + 2\right)^{-1} \left( m^{\alpha_2 + 2}_\mathrm{max} - m^{\alpha_2 + 2}_\mathrm{break} \right) \right], \\
        \left\langle m_2 \right\rangle &= \int dm_1 p(m_1) m_1 \frac{\alpha_q+1}{\alpha_q+2} \frac{1 - \left( \frac{m_{2,\mathrm{min}}}{m_1} \right)^{\alpha_q+2}}{1 -\left( \frac{m_{2,\mathrm{min}}}{m_1} \right)^{\alpha_q+1}}.
    \end{aligned}
\end{equation}

%%%%%%%%%%%%%%%%%%%%%%%%%%%%%%%%%%%%%%%%%%%%%%%%%%%%%%%%%%%%%%%%%%%%%%%%%%%%%%%%%%%%%%%%%%%%%%%%%%%%%%%%%%%%%%%%%%%%%%%%%%%%%%%%%%%%

\section{Binary black hole formation}
\label{app:BBH_formation}

%%%%%%%%%%%%%%%%%%%%%%%%%%%%%%%%%%%%%%%%%%%%%%%%%%%%%%%%%%%%%%%%%%%%%%%%%%%%%%%%%%%%%%%%%%%%%%%%%%%%%%%%%%%%%%%%%%%%%%%%%%%%%%%%%%%%

\subsection{Impact of metallicity}

Metallicity plays a key role in determining the outcome of a massive star life; thus, in the following, we describe a model for the BBH merger rate that explicitly keeps track of the metallicity dependence. 
Starting from equation~\eqref{eq:cosmicSFR_def}, we define the metallicity-dependent star-formation rate density~$\dot{\rho}_\mathrm{SFR}(Z, z)$ as 
\begin{equation}
    \dot{\rho}_\mathrm{CSFR}(z) = \int dZ \dot{\rho}_\mathrm{SFR}(Z, z);
\end{equation}
therefore, the number density of stars formed per year at fixed metallicity is
\begin{equation}
    R_\mathrm{SFR} (Z, z) = \frac{ \dot{\rho}_\mathrm{SFR}(Z, z)}{\left\langle m_\star \right\rangle},
\end{equation}
where~$\left\langle m_\star \right\rangle$ is the average stellar mass of the population.
Only a fraction of these stars ultimately ends up in a binary: for this reason, we introduce a metallicity-dependent efficiency function~$\varepsilon(Z)$ in such a way that the binary formation rate reads as
\begin{equation}
    R_\mathrm{BFR}(Z, z) = \varepsilon(Z) R_\mathrm{SFR} (Z, z).
\end{equation}

From this point on, we need to differentiate between stellar formation and observed merger redshift, labeled~$z_f$ and~$z$, respectively.
Additionally, for the sake of clarity, the metallicity at the stellar formation time is indicated as~$Z_f$.
The causes of a time-delay~$t_d$ between stellar formation and merger depend on the physical effects intervening in different BBH formation mechanisms.
However, we can define, without loss of generality, the merger rate density at redshift~$z$ as
\begin{equation}
    \begin{aligned}
        R_\mathrm{M}(z, Z_f) &= \int_0^{t(z)} dt_d \int dz'_f p(z'_f, t_d| Z_f, z) R_\mathrm{BFR}(Z_f, z'_f) \\
        &= \int_0^{t(z)} dt_d \int dz'_f p(z'_f| t_d, Z_f, z) p(t_d| Z_f, z) R_\mathrm{BFR}(Z_f, z'_f) \\
        &= \int_0^{t(z)} dt_d  \int dz'_f \delta^D(z'_f-z_f(t_d,z)) p(t_d| Z_f) R_\mathrm{BFR}(Z_f, z'_f) \\
        &= \int_0^{t(z)} dt_d  p(t_d| Z_f) R_\mathrm{BFR}(Z_f, z_f(t_d,z)),
    \end{aligned}
\label{eq:merger_rate}
\end{equation}
where~$z_f(t_d,z)$ is the inverse of the function
\begin{equation}
    t_d(z_f,z) = \int_z^{z_f} \frac{dz'}{(1+z')H(z')},
\end{equation}
the upper integration limit for the time-delay is set to~$t(z)$ to select binaries that merge within a Hubble time, $H(z)$ is the Hubble expansion rate, and we assume that the time-delay pdf depends only on local processes, therefore we can drop the redshift dependence.
Given this construction, the total merger rate density is given by
\begin{equation}
    R_\mathrm{M}(z) = \int dZ_f R_\mathrm{M}(z, Z_f).
\end{equation}

\begin{figure}[ht]
    \centerline{
    \includegraphics[width=0.6\columnwidth]{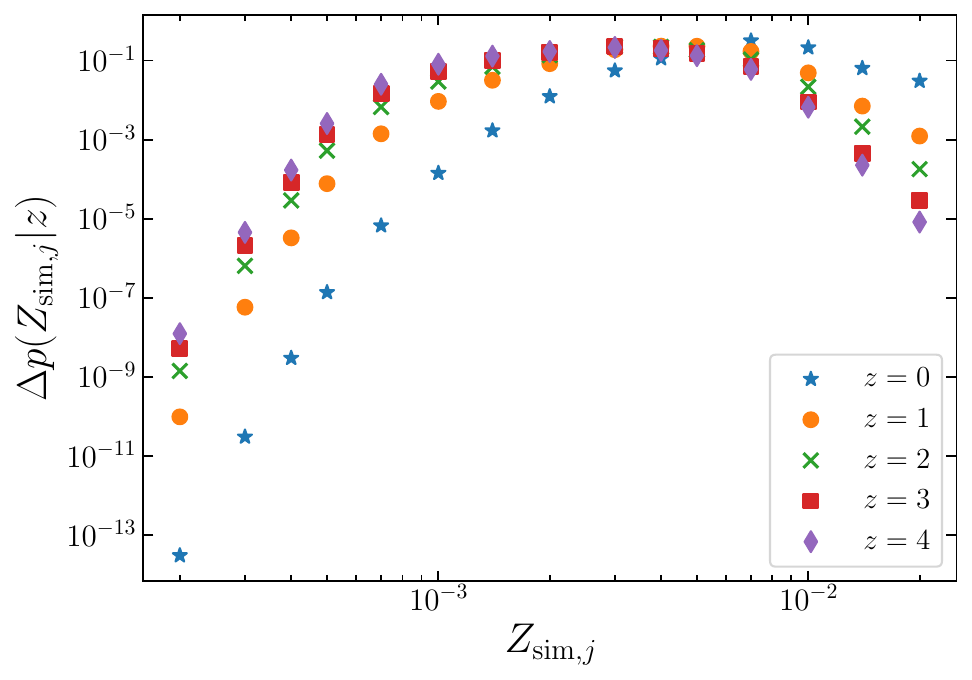}}
    \caption{Metallicity-dependent weights to associate with BBH population simulations at different metallicities.}
\label{fig:metallicity_weights}
\end{figure}

At the practical level, we simulate BBH catalogs for a wide range of metallicity values.
In particular, the analysis of section~\ref{subsec:multi_population_scenario} uses catalogs created for~$15$ different metallicity values
\begin{equation}
    \begin{aligned}
        Z_\mathrm{sim} &= \{ 0.0002, 0.0003, 0.0004, 0.0005, 0.0007, 0.001, 0.0014, 0.002, \\
        &\qquad  0.003, 0.004, 0.005, 0.007, 0.01, 0.014, 0.02 \}.
    \end{aligned}
\end{equation}
Therefore, the properties of the total BBH population will be inferred by weighting the outcome of each simulation at fixed metallicity.
According to appendix~\ref{app:synthetic_universe}, the metallicity PDF is given by
\begin{equation}
    p(Z|z) = \int dM_\star d\mathrm{SFR}  p(Z|M_\star,\mathrm{SFR},z) p(\mathrm{SFR}|M_\star,z) p(M_\star|z),
\end{equation}
independent of the BBH formation mechanism.
Thus, the weight associated with each individual simulation is
\begin{equation}
    \Delta p \left(Z_{\mathrm{sim},j}|z \right) = \int^{\mathcal{Z}_{j+1}}_{\mathcal{Z}_j} dZ p(Z|z),
\end{equation}
where the lower and upper integration limits are~$\mathcal{Z} = \left\lbrace 0.0001, ..., \sqrt{Z_{\mathrm{sim},j+1} Z_{\mathrm{sim},j}}, ... , 0.1 \right\rbrace$.
We show in figure~\ref{fig:metallicity_weights} the weights associated with our synthetic Universe for different redshifts.
We observe a clear progression in which the weight of processes occurring in the high-metallicity tail of the distribution becomes more relevant at low redshift, as expected.

%%%%%%%%%%%%%%%%%%%%%%%%%%%%%%%%%%%%%%%%%%%%%%%%%%%%%%%%%%%%%%%%%%%%%%%%%%%%%%%%%%%%%%%%%%%%%%%%%%%%%%%%%%%%%%%%%%%%%%%%%%%%%%%%%%%%

\subsection{Isolated binary formation channel}

For each value of metallicity, we simulate the evolution of~$N^\mathrm{sim}_\mathrm{bin} = 10^7$ stellar binaries using the binary population-synthesis code~\textsc{SEVN} \cite{spera:sevn, mapelli:sevn, iorio:sevn}.
We adopt the fiducial set-up of Ref.~\cite{iorio:sevn} and follow Ref.~\cite{sgalletta:sfrsb} to generate the initial conditions.
In particular, we assume the common envelope efficiency~$\alpha_\mathrm{CE}=1$, the pair-instability treatment of Ref.~\cite{mapelli:pairinst}, the~\textit{rapid} model for the core-collapse supernovae by Ref.~\cite{fryer:cc} and we sample the natal kicks as in Ref.~\cite{giacobbo:nk}.

In the isolated scenario, the time-delay is given by the sum of two contributions, i.e.,~$t_d = t_\mathrm{BHF} + t_\mathrm{GW}$, where~$t_\mathrm{BHF}$ is the time it takes for both stars to become a BH, and~$t_\mathrm{GW}$ represents the time it takes for the two BHs to merge if energy is radiated exclusively via GWs~\cite{peters:mergingtime}.

\begin{figure}[ht]
    \centerline{
    \includegraphics[width=\columnwidth]{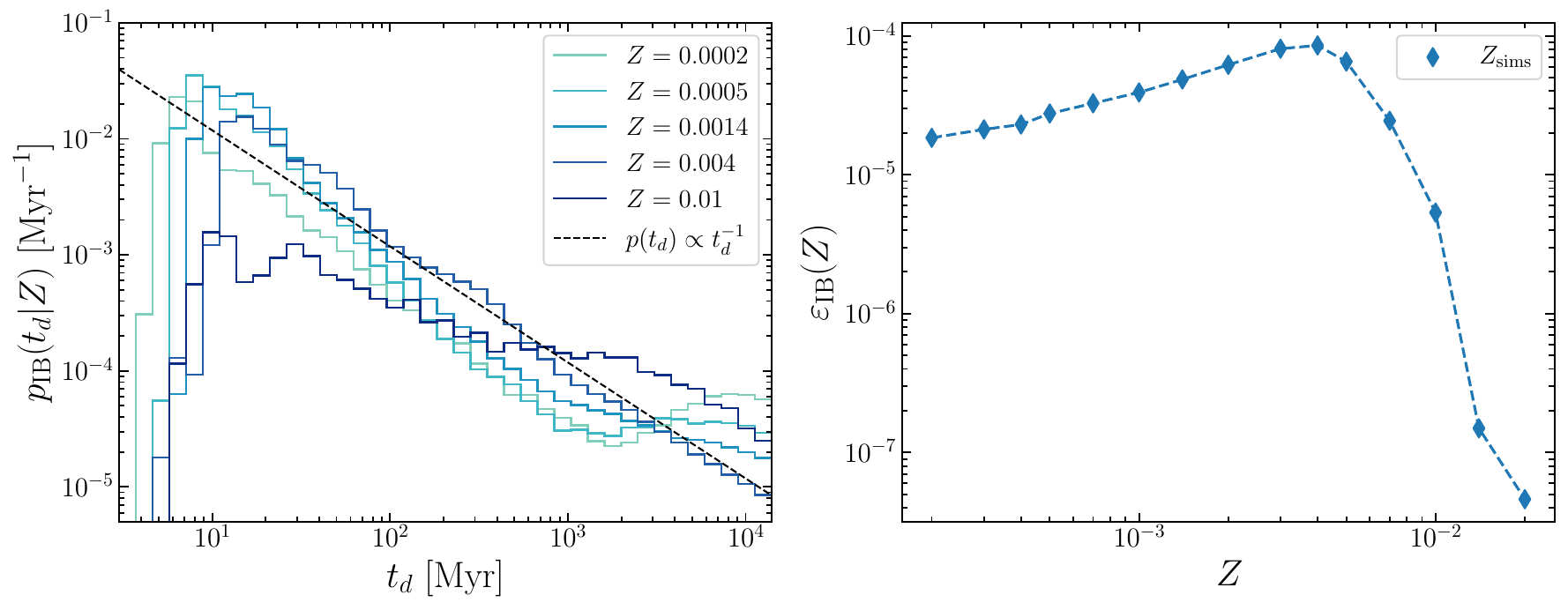}}
    \caption{\textit{Left panel}: time-delay pdf for the isolated binary channel with different environmental metallicity.
    The \textit{black dashed line} represents the commonly used assumption of~$p(t_d) \propto t_d^{-1}$.
    \textit{Right panel}: efficiency of the isolated binary channel in producing BBH that merge in less than a Hubble time as a function of the metallicity.
    \textit{Blue diamonds} represent the value calculated from simulations at fixed metallicity.}
\label{fig:IB_ptd_efficiency}
\end{figure}

The left panel of figure~\ref{fig:IB_ptd_efficiency} shows the time-delay pdfs for five metallicities of our sample.
The formation time of BBHs is approximately uniformly distributed in the time range~$t_\mathrm{BHF} \approx [2,12]\ \mathrm{Myr}$, affecting only the low time-delay tail of the distribution.
Note that time scales of the order of $10\ \mathrm{Myr}$ are negligible compared to cosmological time scales; thus, uncertainties in the formation time have minimal impact on the computation of the GW bias. 
In general, we observe a systematic increment of the time-delay as the metallicity increases.

We define the efficiency for the isolated channel as
\begin{equation}
    \varepsilon_\mathrm{IB}(Z,z) = \frac{N_\mathrm{BBH} \left(Z, t_d\leq t(z) \right)}{N_\star},
\end{equation}
where~$N_\mathrm{BBH} \left(Z, t_d\leq t(z) \right)$ is the number of BBHs that form at a given metallicity and merge within a Hubble time, and~$N_\star$ is the number of stars of the entire stellar population.
However, simulations involve only a subset of the objects that make up the entire binary population.
In fact, only a fraction~$f_\mathrm{bin}$ of stars ends up in binary systems, creating~$N_\mathrm{bin} = (f_\mathrm{bin}/2) N_\star$ binaries.
In this work, we assume an average value of~$f_\mathrm{bin}=0.5$, although massive stars are more likely to be found in binary systems than solar/sub-solar mass stars~\cite{offner:binaryfraction}.
Additionally, only a fraction of such binaries is simulated, i.e., $N^\mathrm{sim}_\mathrm{bin} = f_\mathrm{IMF} N_\mathrm{bin}$.
In practice, we simulate only binaries that have primary mass~$m_1 \geq 5\ M_\odot$ and secondary mass at least~$m_2\geq 2.2\ M_\odot$; therefore, following appendix~\ref{subapp:binary_population}, we have
\begin{equation}
    \begin{aligned}
        f_\mathrm{IMF} &= \int_{5}^{150} dm_1 p(m_1) \int_{\mathrm{max}\left[ 2.2, m_1/10 \right]}^{m_1} dm_2 p(m_2|m_1) \\
        &= \mathcal{N}^{-1}_\star \int_{5}^{150}\!\!\! dm_1 m_1^{\alpha_2} \frac{1 - \left(\mathrm{max}\left[2.2, m_1/10 \right]/m_1 \right)^{\alpha_q+1}}{1-\left(\mathrm{max}\left[0.1, m_1/10 \right]/m_1 \right)^{\alpha_q+1}}.
    \end{aligned}
\end{equation}
Therefore, the efficiency computed from simulation~$\varepsilon^\mathrm{sim}_\mathrm{IB} = N_\mathrm{BBH} \left(Z, t_d\leq t(z) \right) / N^\mathrm{sim}_\mathrm{bin}$ is connected to the theoretical one by~\cite{santoliquido:efficiency}
\begin{equation}
    \varepsilon_\mathrm{IB} = \frac{1}{2} f_\mathrm{IMF} f_\mathrm{bin} \varepsilon^\mathrm{sim}_\mathrm{IB}.
\end{equation}
We report the theoretical value of efficiency as a function of metallicity in the right panel of figure~\ref{fig:IB_ptd_efficiency}.
Our results clearly show how high-metallicity environments effectively inhibit BBH formation, i.e., we expect binary formation to be substantially suppressed at low redshift.

Finally, we end with a technical note. 
In this framework, the time-delay pdf is normalized considering as the maximum time-delay the Hubble time,~$t_0 = t(z=0)$. 
On the other hand, the efficiency already accounts for the fact that some BBH merge in a time larger than Hubble.
This choice does not represent a limitation when moving to redshift~$z\geq 0$, since in reality it is the product of time-delay pdf and efficiency that carries the ``physical meaning'' of representing the fraction of events merging in a given time delay interval.
In other words, since we have
\begin{equation}
	\varepsilon (Z,0) p(t_d\leq t_0|Z) = \varepsilon (Z,z) p(t_d\leq t(z)|Z),
\end{equation}
the reported values can be applied to all redshift.

%%%%%%%%%%%%%%%%%%%%%%%%%%%%%%%%%%%%%%%%%%%%%%%%%%%%%%%%%%%%%%%%%%%%%%%%%%%%%%%%%%%%%%%%%%%%%%%%%%%%%%%%%%%%%%%%%%%%%%%%%%%%%%%%%%%%

\subsection{Globular cluster formation channel}

The second channel explored in this work is the dynamical formation of BBH in globular clusters.
For each metallicity value, we simulate the evolution of~$10^6$ BBHs with the~\textsc{FastCluster} Monte Carlo code~\cite{mapelli:fastclusterI, mapelli:fastclusterII, torniamenti:fastcluster}.
The initial BH population is sampled from the catalogs generated with~\textsc{SEVN} and the BBH dynamical pairing follows the criterion introduced in Ref.~\cite{antonini:pairing}, that tends to couple the most massive objects within the cluster core~\cite{heggie:enc}.
We sample the initial cluster mass from a Gaussian distribution with mean~$\langle\mathrm{log}_{10} M_\mathrm{GC}/M_\odot\rangle=5.9$ and standard deviation~$\sigma_{\mathrm{log}_{10}M_\mathrm{GC}}=0.4$ and the density at half-mass radius is drawn from a Gaussian distribution with mean~$\langle\mathrm{log}_{10} \rho/(M_\odot\mathrm{pc}^{-3})\rangle=4.7$, and standard deviation~$\sigma_{\mathrm{log}_{10}\rho}=0.4$ (see appendix B of Ref.~\cite{torniamenti:fastcluster}).
Following Ref.~\cite{torniamenti:fastcluster}, the BH spin magnitudes are sampled from a Maxwellian distribution with~$\sigma_\chi=0.1$.
Finally, we assume the common envelope efficiency, pair-instability, core-collapse supernovae and natal kick models adopted for the~\textsc{SEVN} runs.

This dynamical scenario allows for the hierarchical formation of multiple binary generations throughout the lifetime of the GC.
We label the properties of first generation binaries with the subscript~``$1\mathrm{g}$'', and the properties of the set of binaries from the second to the $n$-th generation with~``$n\mathrm{g}$''.
Following the formalism adopted by~\textsc{FastCluster}, the time-delay between stellar formation and merger for the first generation of events receives multiple contributions and reads as
\begin{equation}
    t^{(1\mathrm{g})}_d = \mathrm{max}\left[t_\mathrm{BHF}, t_\mathrm{cc} \right] + t_\mathrm{GW},
\end{equation}
where~$t_\mathrm{cc} \approx 3 t_\mathrm{rel}$ is the GC core-collapse time, which represents the typical time it takes for BHs to form, reach the center of the dense GC core and form a binary, and~$t_\mathrm{rel}$ is the half-mass relaxation time.
On the other hand, the time-delay of the second generation is given by
\begin{equation}
    t^{(2\mathrm{g})}_d = t^{(1\mathrm{g})}_d + t_\mathrm{DF} + t_{12} + t_\mathrm{GW},
\end{equation}
where~$t_\mathrm{DF}$ is the typical dynamical friction time necessary for the BH to return to the core, and~$t_{12}$ is the typical dynamical times for the binary exchange formation mechanism.
The reasoning naturally extends to the time-delay of subsequent generations.

However, only a small fraction of events undergo multiple hierarchical mergers.
In terms of relevance, our simulations suggest that the fraction of second generation merging events compared to first generation ones is approximately~$f^{(2\mathrm{g})}(Z) \approx 15-20\%$, whereas higher generation mergers contribute with~$f^{(n\geq 3\mathrm{g})}(Z) \lesssim 1\%$.
Therefore, the global shape of the time-delay pdf is strongly dominated by first generation binaries.
The presence of multiple generations of events requires a slight modification of the formalism introduced above.
Since the relative contribution of merger from generations above the second is negligible, we consider only the first two.
In practice, we define the first and second generation merger rate as
\begin{equation}
    \begin{aligned}
        R^{(1\mathrm{g})}_M(z) &= \int_0^{t(z)} dt_d  p^{(1\mathrm{g})}(t_d| Z_f) \varepsilon^{(1\mathrm{g})}_\mathrm{GC}(Z_f, z_f) R_\mathrm{SFR}(Z_f, z_f(t_d,z)), \\
        R^{(2\mathrm{g})}_M(z) &= \int_0^{t(z)} dt_d  p^{(2\mathrm{g})}(t_d| Z_f) f^{(2\mathrm{g})}(Z_f) \varepsilon^{(1\mathrm{g})}_\mathrm{GC}(Z_f, z_f) R_\mathrm{SFR}(Z_f, z_f(t_d,z)),
    \end{aligned}
\end{equation}
in such a way that the total merger rate is~$R_M(z) = R^{(1\mathrm{g})}_M(z) + R^{(2\mathrm{g})}_M(z)$.
In this sense, the only efficiency we need to estimate is that of first generation mergers.
Assuming that we have one binary per GC, we have
\begin{equation}
    \varepsilon^{(1\mathrm{g})}_\mathrm{GC}(z) = \frac{f^\mathrm{GC}_\mathrm{SFR}(z)}{\left\langle N_\mathrm{star/GC}(z) \right\rangle},
\end{equation}
where the fraction of star-formation occurring in GC is~$f^\mathrm{GC}_\mathrm{SFR} = \dot{\rho}_\mathrm{SFR}/\dot{\rho}_\mathrm{CSFR}$, the GC star-formation rate density is provided in Ref.~\cite{elbadry:globularclustersimulation}, the average number of stars per GC is
\begin{equation}
    \left\langle N_\mathrm{star/GC}(z) \right\rangle = \frac{\left\langle M_\mathrm{GC}(z) \right\rangle}{\left\langle m_\star \right\rangle} = \eta \frac{\left\langle M_h(z) \right\rangle}{\left\langle m_\star \right\rangle},
\end{equation}
and~$\left\langle M_h(z) \right\rangle$ is the average halo mass.
Given our assumptions, the GC efficiency turns out to be independent of metallicity.

\begin{figure}[ht]
    \centerline{
    \includegraphics[width=\columnwidth]{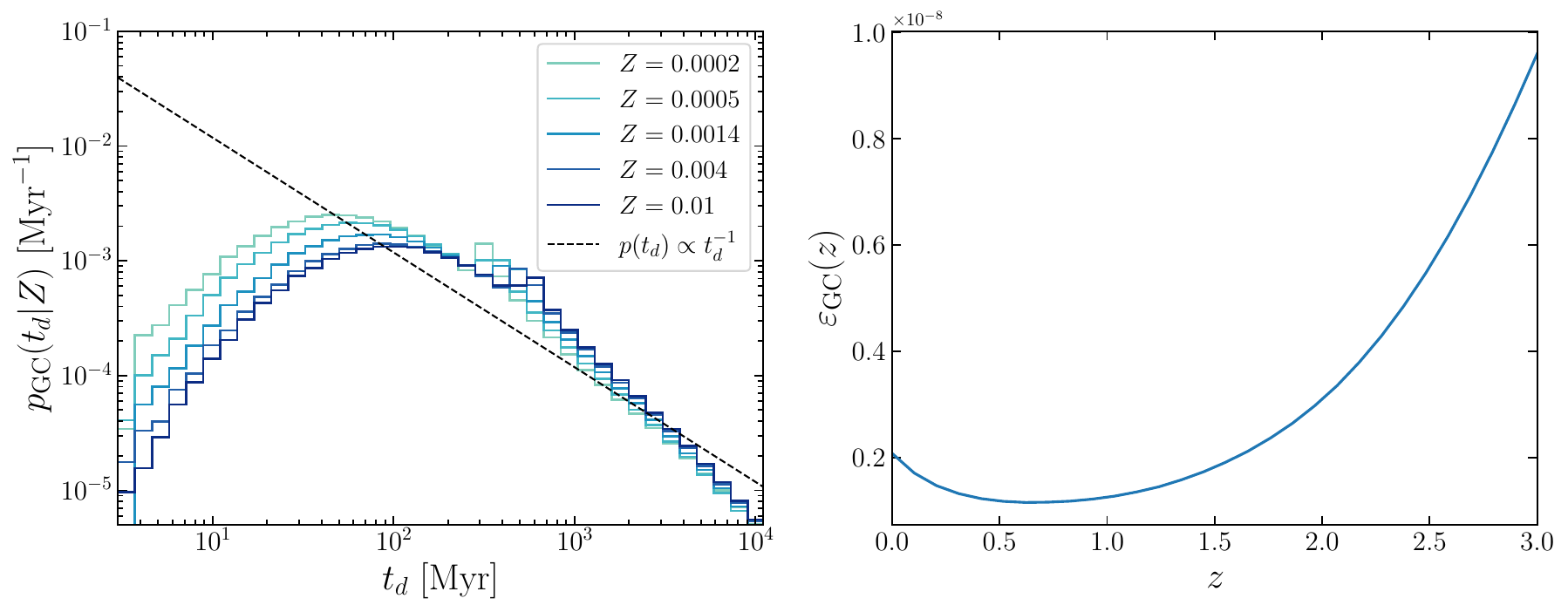}}
    \caption{\textit{Left panel}: time-delay pdf for the GC channel with different environmental metallicity.
    The \textit{black dashed line} represents the commonly used assumption of~$p(t_d) \propto t_d^{-1}$.
    \textit{Right panel}: efficiency of the GC channel in producing BBH that merge in less than a Hubble time as a function of the redshift.}
\label{fig:GC_ptd_efficiency}
\end{figure}

The left panel of figure~\ref{fig:GC_ptd_efficiency} shows the time-delay distribution that includes all generations of mergers.
In contrast to the isolated binary scenario, here we observe s rather sharp deviation from the power-law distribution, although in this case metallicity appears to play a minor role.
On the other hand, in the right panel of the same figure we observe that the efficiency of the process is lower than in the isolated binary scenario since GC contribution to star-formation rate is rather small, with~$f^\mathrm{GC}_\mathrm{SFR} \approx 1-3\times 10^{-3}$ in the late Universe.

%%%%%%%%%%%%%%%%%%%%%%%%%%%%%%%%%%%%%%%%%%%%%%%%%%%%%%%%%%%%%%%%%%%%%%%%%%%%%%%%%%%%%%%%%%%%%%%%%%%%%%%%%%%%%%%%%%%%%%%%%%%%%%%%%%%%

\subsection{Sampling the properties of binary black hole hosts}

The catalogs obtained in the previous sections are informative exclusively about the binary properties. 
Therefore, the final step in studying the GW clustering properties requires sampling the properties of the hosts.
Regarding the analysis of section~\ref{subsec:model_agnostic_estimate}, the host masses at both formation and merger redshift are estimated as explained in Ref.~\cite{bellomo:classgwb}.
On the other hand, the model used in section~\ref{subsec:multi_population_scenario} requires a more advanced approach to ensure compatibility with the synthetic Universe approach developed in appendix~\ref{app:synthetic_universe}.

In particular, in that scenario we have the necessity of sampling from multidimensional pdfs.
As in the 1D case, also in multiple dimensions we can apply recursively the inverse transform sampling method.
For instance, let us consider the case of a 2D pdf~$p_\mathrm{2D}(x,y)$, where~$(x,y)\in [x_\mathrm{min},x_\mathrm{max}]\times[y_\mathrm{min},y_\mathrm{max}]$.
A sample of the first variable is obtained as~$x=c^{-1}_\mathrm{1D}(u_x)$, where~$u_x\sim\mathcal{U}(0,1)$ and we introduced the 1D cumulative distribution function
\begin{equation}
	c_\mathrm{1D}(x) = \int_{x_\mathrm{min}}^{x} dx'\ p_\mathrm{1D}(x') = \int_{x_\mathrm{min}}^{x} dx' \int_{y_\mathrm{min}}^{y_\mathrm{max}} dy\ p_\mathrm{2D}(x',y) = u_x.
\end{equation}
Once~$x$ is sampled, we obtain~$y$ by inverting
\begin{equation}
	c_\mathrm{1D}(y|x) = \int_{y_\mathrm{min}}^{y} dy' p_\mathrm{1D}(y'|x) = \int_{y_\mathrm{min}}^{y} dy' \frac{p_\mathrm{2D}(x,y')}{p_\mathrm{1D}(x)} = u_y,
\end{equation}
i.e.,~$y=c^{-1}_\mathrm{1D}(u_y|x)$ with~$u_y\sim\mathcal{U}(0,1)$.
In this fashion, the~$(x,y)$ pair has been effectively sampled from~$p_\mathrm{1D}(y|x)p_\mathrm{1D}(x) \equiv p_\mathrm{2D}(x,y)$, as desired.
The procedure extends to an arbitrary number of random variables since the conditional probability law can be applied recursively.

In the IB case, we sample the merger redshift from the observed merger rate pdf~$p(z) \propto (1+z)^{-1}R_\mathrm{M} dV/dz$, the time-delay and metallicity at formation from the 2D pdf~$$p(t_d,Z_f) \propto p(t_d| Z_f) \varepsilon(Z_f) R_\mathrm{SFR} (Z_f, z_f),$$ the SFR and stellar mass of the host galaxy from~$p(\mathrm{SFR}, M_\star | z_f)$, and the halo mass at formation from a pdf~$p(M_h|M_\star,z_f) \propto d^2n_\mathrm{gal}/dM_hdM_\star$ at fixed stellar mass.
The halo mass at merger is computed by evolving the halo forward in time down to the merger redshift, as explained in Ref.~\cite{bellomo:classgwb}.
On the other hand, in the GC case, we sample merger redshift, time-delay and metallicity at formation as reported above, the GC mass from a pdf~$p(M_\mathrm{GC}) \propto dn_\mathrm{GC}/dM_\mathrm{GC}$, and the halo mass at formation from~$p(M_h|M_\mathrm{GC}, z_f) \propto d^2n_\mathrm{GC}/dM_hdM_\mathrm{GC}$ at fixed GC mass.
The halo mass at merger is computed as for the IB case.